\documentclass[aps,prd,twocolumn,superscriptaddress,preprintnumbers,floatfix,nofootinbib,notitlepage,showkeys,showpacs, amsmath,amssymb]{revtex4-2}
 
\usepackage{graphicx}
\usepackage{dcolumn}
\usepackage{bm}
\usepackage{hyperref}
\usepackage{xcolor}

\usepackage{simpler-wick}
\usepackage{simplewick}
\usepackage[normalem]{ulem}

\usepackage{scalerel,tikz}
\usetikzlibrary{svg.path}
\definecolor{orcidlogocol}{HTML}{A6CE39}
\tikzset{orcidlogo/.pic={
 \fill[orcidlogocol] svg{M256,128c0,70.7-57.3,128-128,128C57.3,256,0,198.7,0,128C0,57.3,57.3,0,128,0C198.7,0,256,57.3,256,128z};
 \fill[white] svg{M86.3,186.2H70.9V79.1h15.4v48.4V186.2z}
 svg{M108.9,79.1h41.6c39.6,0,57,28.3,57,53.6c0,27.5-21.5,53.6-56.8,53.6h-41.8V79.1z M124.3,172.4h24.5c34.9,0,42.9-26.5,42.9-39.7c0-21.5-13.7-39.7-43.7-39.7h-23.7V172.4z}
 svg{M88.7,56.8c0,5.5-4.5,10.1-10.1,10.1c-5.6,0-10.1-4.6-10.1-10.1c0-5.6,4.5-10.1,10.1-10.1C84.2,46.7,88.7,51.3,88.7,56.8z};
}}
\newcommand\orcidicon[1]{\href{https://orcid.org/#1}{\mbox{\scalerel*{
\begin{tikzpicture}[yscale=-1,transform shape]
\pic{orcidlogo};
\end{tikzpicture}
}{|}}}}

\begin{document}

\preprint{APS/123-QED}

\title{Delensing CMB $B$-modes using galaxy surveys: \\ the effect of galaxy bias and matter clustering non-linearities}
\author{Shengzhu Wang~\orcidicon{0009-0008-0142-1794}}
\affiliation{Berkeley Center for Cosmological Physics, UC Berkeley, CA 94720, USA}
\affiliation{Department of Physics, University of California, Berkeley, CA 94720, USA}
\author{Ant\'on Baleato Lizancos~\orcidicon{0000-0002-0232-6480}}
\affiliation{Berkeley Center for Cosmological Physics, UC Berkeley, CA 94720, USA}
\affiliation{Department of Physics, University of California, Berkeley, CA 94720, USA}
\affiliation{Lawrence Berkeley National Laboratory, One Cyclotron Road, Berkeley, CA 94720, USA}
\author{Jos\'e Luis Bernal~\orcidicon{0000-0002-0961-4653}}
\affiliation{Instituto de Física de Cantabria (IFCA), CSIC-Univ. de Cantabria, Avda. de los Castros s/n, E-39005 Santander, Spain}

\begin{abstract}
The $B$-mode of polarization of the CMB is a uniquely powerful probe of gravitational waves produced in the very early Universe. But searches for primordial $B$-mode anisotropies must contend with gravitational lensing, which induces late-time $B$-modes not associated with gravitational waves. These lensing $B$-modes can be removed --- i.e., `delensed' --- using observations of the $E$-modes and a proxy of the matter fluctuations along the line of sight that caused the deflections. The number density and redshift reach of galaxy surveys such as the upcoming Rubin observatory offer attractive prospects for using them to delens $B$-mode data from CMB experiments such as the Simons Observatory, LiteBIRD or CMB-S4. However, stochasticity and non-linear galaxy bias may in principle decorrelate the galaxy field from the matter distribution responsible for the lensing effect, thus hindering efforts to delens $B$-modes. In addition, non-linear gravitational evolution and bias  introduce non-Gaussianities in the large-scale structure which further complicate the modelling. We quantify these effects by populating an $N$-body simulation with a magnitude-limited, photometric sample of galaxies similar to Rubin's \emph{gold} selection, and using them to delens CMB maps lensed by the same matter distribution. We find that pipelines that treat the galaxy overdensity as a Gaussian field will incur negligible bias on the inferred tensor-to-scalar ratio, $r$. Moreover, we show that even in a highly conservative scenario where only the linear bias of the galaxies can be determined, the bias on $r$ arising from this simplification is well within statistical uncertainties for a cosmic-variance limited scenario where Rubin-like galaxies are used for delensing.
\end{abstract}

\maketitle 


\section{Introduction}
The cosmic microwave background (CMB) contains a wealth of information about our Universe. While `secondary' anisotropies such as lensing~\cite{ref:lewis_challinor_review} or the Sunyaev Zel'dovich effects~\cite{carlstrom_cosmology_2002} tell us about the distribution and properties of baryonic and dark matter at late times, the `primary' anisotropies constitute a window onto conditions around the recombination era, approximately 380,000 years after the Big Bang. Moreover, primordial gravitational waves produced by inflationary expansion in the very early Universe~\cite{ref:guth_81, ref:linde_82, ref:albrecht_steinhardt_82, ref:starobinsky_79} may have left their signatures on the CMB~\cite{Polnarev:1985}. The cleanest observable to detect this
phenomenon is the anistropies in the $B$-mode of polarization of the CMB, because (to leading order) the primordial $B$-mode does not receive contributions from scalar (density) perturbations and is thus free of cosmic variance~\cite{Seljak:1996gy, Kamionkowski:1996zd}. For this reason, the search for CMB $B$-modes has become a key pursuit for cosmologists. 

Since there is no lower limit to the amplitude of the tensor modes, progress in this endeavor requires exquisite sensitivity and control of systematics, particularly regarding foreground contamination (see, e.g., Ref.~\cite{ref:kamionkowski_kovetz_21}). Another major challenge stems from the fact that gravitational lensing converts $E$-modes into $B$-modes, and these lensing-induced $B$-modes act as as source of noise when searching for the primordial component~\cite{ref:zaldarriaga_seljak_98}. In order to reduce the  
sample variance related with this contribution, an estimate of the map-level realization of the lensing $B$-modes must be generated and removed, a procedure known as `delensing' that has already been studied extensively~\cite{ref:knox_2002, Seljak:2003pn, ref:carron_19, ref:millea_et_al_20, ref:SO_delensing_paper, ref:bharat_forecasts, ref:diego-palazuelos_et_al_20, ref:baleato_20_internal, ref:cib_delensing_biases, ref:S4_pgw_forecasts_22, ref:BK_moriond_22, baleato_lizancos_impact_2022, aurlien_foreground_2023,heinrich_cmb_2022,belkner_cmb-s4_2023, floss_denoising_2024,trendafilova_improving_2023, hertig_simons_2024} and successfully applied to data~\cite{ref:carron_17, ref:manzotti_delensing, Planck2018:lensing, ref:polarbear_delensing_19, ref:han_20, ref:bicep_delensing}.

All that is needed to produce and remove these maps of the lensing $B$-modes are measurements of the $E$-modes and some estimate of the lensing potential. The latter can be obtained either internally through CMB lensing reconstructions (see e.g.,Ref.~\cite{ref:lewis_challinor_review}), or externally by invoking other tracers of the matter distribution, such as the cosmic infrared background~\cite{ref:sherwin_15} or maps of the galaxy distribution~\cite{ref:smith_12_external, ref:manzotti_18}.

In recent years, several works have demonstrated the robustness of external delensing of $B$-modes on real data~\cite{ref:bicep_delensing, ref:manzotti_delensing, hertig_25}, and even used it to improve constraints on the tensor-to-scalar ratio, $r$~\cite{ref:bicep_delensing}. This approach will be an important component of the delensing program for upcoming experiments such as the Simons Observatory (SO)~\cite{ref:SO_science_paper}, LiteBIRD~\cite{ref:litebird}, and any other experiment measuring large-scale $B$-modes in regions of the sky where ultra-deep, high-resolution observations of the
CMB polarization are not available (the latter are a requirement for the ultimately-optimal, internal delensing of $B$-modes; see, e.g., Refs.~\cite{ref:SO_delensing_paper, ref:bharat_forecasts, ref:diego-palazuelos_et_al_20, ref:S4_pgw_forecasts_22, ref:BK_moriond_22}). `External delensing' may prove important even for CMB-S4, either as a way of boosting the delensing achieved through internal methods, or as an external cross-check.

So far, the potentiality of external tracers to yield extensive delensing has been established in highly idealized scenarios, with only a few works studying the impact of systematics (see e.g., Refs.~\cite{ref:namikawa_16, ref:cib_delensing_biases}). An area of particular concern
yet to be studied is the impact of non-linearities in structure formation and galaxy bias, as well as galaxy stochasticity. Non-Gaussianities in the galaxy distribution can couple with non-Gaussianities in the lensing potential and give rise to contributions to the power spectrum of delensed $B$-modes that are not captured by typical models, which assume Gaussianity. In addition, galaxy bias can decorrelate the observed galaxy field from the gravitational lensing potential, hindering the delensing efficiency and potentially further complicating the modeling. Reference~\cite{ref:namikawa_and_takahashi_19} were the first to study the impact of non-linear matter clustering perturbations on $B$-mode delensing, finding a negligible effect. Here, we extend their work by additionally characterizing the impact of non-linear galaxy bias and stochasticity.

To study the impact of these effects on $B$-mode delensing, we populate an $N$-body simulation with galaxies resembling the magnitude-limited \emph{gold} sample of Rubin and use it to delens CMB maps which have been lensed by that same underlying matter distribution. We then compare the delensing performance using this `realistic' galaxy sample to more standard and simplistic pipelines that assume Gaussianity and linear galaxy bias. These simplifications lead to very small corrections to the power spectrum of delensed $B$-modes --- smaller than cosmic variance on a per-mode basis --- and show that these translate to biases on the tensor-to-scalar ratio which are smaller than the 1-$\sigma$ statistical uncertainty expected of upcoming experiments such as SO. In addition, our results suggest that idealized forecasts of delensing performance that assume galaxy bias is strictly linear are actually rather accurate, with corrections being smaller than 5\% per mode (for these Rubin galaxies). Our results demonstrate the robustness of delensing $B$-modes with external tracers and warrant this approach to boost the sensitivity to primordial tensor modes.

The structure of this paper is as follows. We start with a theoretical description of CMB lensing and delensing in section~\ref{sec:theory_intro}. We then describe our simulations in section~\ref{sec:simulations}. Our main results are presented in section~\ref{sec:results}. Discussions and conclusions can be found in section~\ref{sec:conclusions}. Additional technical details are relegated to the appendices.

Throughout this work, we assume a flat $\Lambda$CDM cosmology consistent with Ref.~\cite{ref:planck_15_params}, as assumed in the  MultiDark Planck 2 (MDPL2) simulation~\cite{Klypin:2014kpa, Rodriguez-Puebla:2016ofw}, which is the underlying $N$-body simulation employed throughout this work.

\section{Theory of CMB Lensing and Delensing}\label{sec:theory_intro}
\subsection{CMB lensing}
Gravitational lensing remaps the CMB anisotropies at some intrinsic angular position $\boldsymbol{\hat{n}}$ on the sky onto the observed angular position $\boldsymbol{\hat{n}}'$ (see e.g., Ref.~\cite{ref:lewis_challinor_review} for a review). The two are related by
\begin{eqnarray}
\boldsymbol{\hat{n}}' = \boldsymbol{\hat{n}}+\boldsymbol{\alpha}(\boldsymbol{\hat{n}}),
\label{deflection_angle}
\end{eqnarray}
where ${\boldsymbol \alpha}$ is the deflection angle. This means that the observed polarization field $\tilde{P}$ can be expressed in terms of the intrinsic one as
\begin{eqnarray}
\tilde{P}(\boldsymbol{\hat{n}})=P(\boldsymbol{\hat{n}}+\boldsymbol{\alpha}(\boldsymbol{\hat{n}}))\,.
\label{lensing_pol}
\end{eqnarray}
Throughout this work, tildes will mark lensed fields.

At leading order, the deflection angle can be fully described in terms of a scalar lensing potential, $\phi$, as
\begin{eqnarray}
\boldsymbol{\alpha}(\boldsymbol{\hat{n}})=\boldsymbol{\nabla} \phi(\boldsymbol{\hat{n}})\,.
\label{lensing_potential}
\end{eqnarray}
This is related to the frequently-used lensing convergence $\kappa$ as
\begin{eqnarray}
\kappa = -\frac{1}{2}\nabla^{2}\phi \,.
\label{lensing_convergence}
\end{eqnarray}
The convergence field at some sky position $\hat{\bm{n}}$ is related to the 3D matter overdensity $\delta_m$ as
\begin{equation}\label{eqn:kappa_ito_delta}
    \kappa (\hat{\bm{n}}) = \int d\chi W^{\kappa}(\chi) \delta_m(\hat{\bm{n}}\chi)\,,
\end{equation}
with the CMB lensing kernel defined as
\begin{equation}
     W^{\kappa}(\chi)\equiv \frac{3}{2} \Omega_m^2 H_0^2 (1+z) \frac{\chi(\chi_*-\chi)}{\chi_*}\,,
\end{equation}
in a flat cosmology, with $\chi_*$ being the comoving distance to the surface of last scattering. These expressions show that CMB lensing is sourced by the total matter  distribution.

It is well known that the distortion of the polarization field due to gravitational lensing shown in equation~\eqref{lensing_pol} converts primordial $E$-modes into $B$-modes. To leading order, and in the flat-sky limit, this can be written in harmonic space as (see e.g.,~\cite{ref:limitations_paper})
\begin{eqnarray}
\tilde{B}(\bm{\ell})=\int \frac{d^{2}\bm{\ell}'}{(2\pi)^2}W(\bm{\ell},\bm{\ell}')E(\bm{\ell}')\kappa(\bm{\ell}-\bm{\ell}')\,,
\label{eqn:flat-sky_lensed}
\end{eqnarray}
with 
\begin{eqnarray}
W(\bm{\ell},\bm{\ell}')=\frac{2\bm{\ell}'\cdot(\bm{\ell}-\bm{\ell}')}{|\bm{\ell}-\bm{\ell}'|^{2}}\sin(2\psi_{\bm{\ell},\bm{\ell}'}),
\label{w_filter}
\end{eqnarray}
where $\psi_{\bm{\ell},\bm{\ell}'}$ is the angle between 2D Fourier-space vectors $\bm{\ell}$ and $ \bm{\ell}'$. Note that we have assumed that there are no primordial $B$-modes.

Here and throughout the body of the paper we will provide flat-sky expressions for readability. Details of the harmonic-space equivalents are provided in  appendix~\ref{appendix_curved}; we will use those to produce our main numerical results.

\subsection{Template-delensing of CMB $B$-modes}
To estimate the lensing $B$-mode realization present on the sky and mitigate its sample variance --- i.e., to delens --- we follow the standard approach of constructing a lensing $B$-mode `template' and subtracting it from observations.

We use a `gradient'-order template, as this has been shown to be effectively optimal for all foreseeable applications~\cite{ref:limitations_paper}. This template is built following equation~\eqref{eqn:flat-sky_lensed}, but replacing the unlensed (or delensed) $E$-modes with the lensed ones to obtain improved delensing performance by cancelling out higher-order contributions~\cite{ref:limitations_paper}. We will assume throughout that we have access to noiseless observations of the $E$- and $B$-modes, as we are interested in isolating $B$-mode residuals due to mismodeling.\footnote{Noise in the $B$-modes will not affect the performance of delensing $B$-modes nor the significance of the impact due to non linearities; hence, it fully decouples from the problem we aim to study in this work. On the other hand, while noise in the $E$-modes can in principle affect the delensing efficiency, it will be a small effect for upcoming experiments. Since it is also highly dependent on the specific experimental configuration, we leave it to specific experiments to study its effect.} Given some proxy of the lensing convergence, $\kappa^{\rm proxy}$, which may for example be a map of the galaxy number-density fluctuations or the cosmic infrared background  anisotropies, the template can be constructed as
\begin{align}
\tilde{B}^{\rm temp}(\bm{\ell})=\int & \frac{d^{2}\bm{\ell}'}{(2\pi)^2}W(\bm{\ell},\bm{\ell}')f(|\bm{\ell}-\bm{\ell}'|)  \tilde{E}(\bm{\ell}')\kappa^{\rm proxy}(\bm{\ell}-\bm{\ell}') \nonumber\\
& \equiv \mathcal{B} \left[ \tilde{E}, \kappa^{\rm proxy}\right] (\bm{\ell})\,,
\label{flat-sky_template}
\end{align}
where we have defined $\mathcal{B}[\tilde{E}, \kappa^{\rm proxy}]$ as a functional to explicitly indicate the dependence of the $\tilde{B}^{\rm temp}$ on the observed $E$-mode fluctuations and the large-scale-structure tracer used as proxy of the convergence field. The weighting function
\begin{eqnarray}
f(\bm{\ell})=\frac{C_{\ell}^{\kappa\kappa^{\rm proxy}}}{C_{\ell}^{\kappa^{\rm proxy}\kappa^{\rm proxy}}}
\label{f_weighting}
\end{eqnarray}
is designed to minimize the variance of the power spectrum of delensed $B$-modes~\cite{ref:smith_12_external}. The angular power spectra in equation~\eqref{f_weighting}
can be determined by fitting a smooth, empirical model to the data; we explain the functional forms we use to do this in appendix~\ref{appendix:power_spectrum}.

Note that $\kappa^{\rm proxy}$ can actually be a combination of several tracers. In order to improve delensing, it is possible to use several tomographic subsamples for the tracers and employ the
optimal weights derived in Ref.~\cite{ref:sherwin_15}. One must then replace $f \kappa^{\rm proxy}$ in equation~\eqref{flat-sky_template} with an optimally-weighted, combined proxy. When galaxy catalogs are used as a proxy for the lensing convergence, $f\kappa^{\rm proxy}$ should be substituted with $\sum_{i}c_{i}\delta_{\rm g}^{i}$, where $\delta_{\rm g}^i$ is the $i$-th sample of the galaxy number overdensity field, and the weights $c_i$ are given by
\begin{eqnarray}
c_{i}=\sum_{j}([\rho^{gg}_\ell]^{-1})^{ij}\rho_\ell^{g_j \kappa}\sqrt{\frac{C_{\ell}^{\kappa g_{i}}}{C_{\ell}^{g_{i}g_{i}}}}\,,
\label{combine_weights}
\end{eqnarray}
where
\begin{eqnarray}
\rho_{\ell}^{ab}=\frac{C_{\ell}^{ab}}{\sqrt{C_{\ell}^{aa}C_{\ell}^{bb}}}\,,
\label{correlation_coeff}
\end{eqnarray}
is the cross-correlation coefficient between the fields $a$ and $b$. In the equation above, 
$([\rho^{gg}_\ell]^{-1})^{ij}$ is the $ij$ element of the inverse of the matrix composed of all correlation coefficients between the $i$-th and the $j$-th galaxy density samples, and $\rho_\ell^{g_j \kappa}$ is the correlation coefficient between the $j$-th galaxy sample and the lensing convergence. Note that the superscript `$g$' in the power spectra in equation~\eqref{combine_weights} refers to the galaxy overdensity field. The generalization of equation~\eqref{combine_weights} to any combination of large-scale structure (LSS) tracers is straightforward.

Let us return to the case of a single sample to delens (from which the multi-sample prediction can be obtained trivially) to ease notation. In this case, the residual lensing $B$-modes after delensing are given by
\begin{eqnarray}
B^{\rm res}(\bm{\ell})=\tilde{B}(\bm{\ell})-\tilde{B}^{\rm temp}(\bm{\ell})=\int \frac{d^{2}\bm{\ell}'}{(2\pi)^2}W(\bm{\ell},\bm{\ell}')\nonumber\\
\times[E(\bm{\ell}')\kappa(\bm{\ell}-\bm{\ell}')-f(|\bm{\ell}-\bm{\ell}'|)\tilde{E}(\bm{\ell}')\kappa^{\rm proxy}(\bm{\ell}-\bm{\ell}')]\,.
\label{residual_$B$-mode_flat}
\end{eqnarray}

Let us write out explicitly the terms that make up the angular power spectrum of residual lensing $B$-modes after delensing:\footnote{Once again, we ignore contributions from experimental noise, foregrounds, and other complications which are tangential to the questions we are addressing.}
\begin{equation}
    C_{\ell}^{BB, \mathrm{res}} \equiv C_{\ell}^{\tilde{B}\tilde{B}} - 2\,C_{\ell}^{\tilde{B}  \tilde{B}^{\rm temp}} + C_{\ell}^{\tilde{B}^{\rm temp}  \tilde{B}^{\rm temp}}\,,
\label{eq:lensing_residual_def}
\end{equation}
where $C_{\ell}^{\tilde{B}\tilde{B}}$ is the power spectrum of lensed $B$-modes, and the cross-spectrum between the true lensed $B$-modes and the template is given by
\begin{align}
C_{\ell}^{\tilde{B} \tilde{B}^{\rm temp}} & = \langle \tilde{B}(\bm{\ell}) \tilde{B}^{\rm temp}(\bm{\ell}') \rangle' \nonumber \\
& = \langle \tilde{B} (\bm{\ell}) \mathcal{B}\left[\tilde{E}, \kappa^{\rm proxy}\right](\bm{\ell}') \rangle' \, .
\end{align}
The prime following the angle bracket in the equation above denotes the Dirac delta function $\delta(\bm{\ell}-\bm{\ell}')$, which has been removed to simplify notation. We have also written the auto-spectrum of the template as
\begin{align}
    C_{\ell}^{\tilde{B}^{\rm temp}  \tilde{B}^{\rm temp}} & = \langle \tilde{B}^{\rm temp}(\bm{\ell}) \tilde{B}^{\rm temp}(\bm{\ell}') \rangle' \nonumber \\
    &= \int \frac{d^2\bm{\ell}_1}{(2\pi)^2} W(\bm{\ell},\bm{\ell}_1) f(|\bm{\ell}-\bm{\ell}_1|) \nonumber \\
    &\quad\times \int \frac{d^2\bm{\ell}_2}{(2\pi)^2} W(\bm{\ell}',\bm{\ell}_2)  f(|\bm{\ell}'-\bm{\ell}_2|) \nonumber\\
    & \qquad \times \langle \tilde{E}(\bm{\ell}_1)  \kappa^{\rm proxy} (\bm{\ell} - \bm{\ell}_1) \nonumber\\
    & \quad\qquad\times \tilde{E}(\bm{\ell}_2) \kappa^{\rm proxy}(\bm{\ell}' - \bm{\ell}_2)\rangle \nonumber' \\
    & = \langle \mathcal{B}\left[\tilde{E}, \kappa^{\rm proxy}\right](\bm{\ell}) \mathcal{B}\left[\tilde{E}, \kappa^{\rm proxy}\right](\bm{\ell}') \rangle' \, .
\end{align}

If $\kappa^{\rm proxy}$ is Gaussian-distributed, it has vanishing connected moments beyond the second one, and the power spectrum of residual $B$-modes defined in equation~\eqref{eq:lensing_residual_def} reduces to
\begin{align}\label{eqn:full_model}
C_{\ell}^{BB, \rm res}\approx\int & \frac{d^{2}\bm{\ell}'}{(2\pi)^2}W^{2}(\bm{\ell},\bm{\ell}')C_{\ell'}^{EE}\nonumber\\
& \times \big[ C_{|\bm{\ell}-{\bm{\ell}'}|}^{\kappa\kappa} - 2 f_{|\bm{\ell}-{\bm{\ell}'}|}C_{|\bm{\ell}-{\bm{\ell}'}|}^{\kappa \kappa^{\rm proxy}} \nonumber \\
& \quad \quad + f^2_{|\bm{\ell}-{\bm{\ell}'}|}C_{|\bm{\ell}-{\bm{\ell}'}|}^{\kappa^{\rm proxy}\kappa^{\rm proxy}} \big]\,.
\end{align}
In the limit that the fiducial spectra used to determine the galaxy weights $f(\pmb{\ell})$ match the truth, we recover the more intuitive result:
\begin{eqnarray}
C_{\ell}^{BB, \rm res}=\int \frac{d^{2}\bm{\ell}'}{(2\pi)^2}W^{2}(\bm{\ell},\bm{\ell}')C_{\ell'}^{EE}C_{|\bm{\ell}-{\bm{\ell}'}|}^{\kappa\kappa}\nonumber\\
\times (1-[\rho^{\kappa\kappa_{\rm proxy}}_{|\bm{\ell}-\bm{\ell}'|}]^2)\,.
\label{residual_power_spectrum_flat}
\end{eqnarray}
Whenever we use these models, we will obtain the cross-correlation coefficients either directly from measurements from the simulations, or by fitting these with the theoretical fitting forms for angular power spectra described in appendix~\ref{appendix:power_spectrum}. 

As mentioned earlier, the power spectra used to determine the weights in equations~\eqref{f_weighting} or~\eqref{combine_weights} can be obtained from smooth fits to the data. Any inaccuracies in their determination are expected to result in suboptimal delensing,\footnote{Delensing performance may be degraded when the fiducial spectra from which the filters are calculated do \emph{not} match the truth. Since the variance of the $B$-mode power spectrum after delensing scales with the residual lensing $B$-mode power, the significance we report in this work for biases on $r$ can be regarded as upper bounds.} but no bias, since the chosen filters are known exactly by construction and can be folded into the model in equation~\eqref{residual_power_spectrum_flat}~\cite{ref:yu_17, ref:SO_delensing_paper, baleato_lizancos_model_2023}. 

The goal of delensing becomes manifest when we consider the per-multipole uncertainty on the measurement of the power spectrum of $B$-modes after delensing. The dominant, disconnected component is
\begin{equation}
\sigma_{\rm res}(\ell)=\sqrt{\frac{2}{(2\ell+1)f_{\rm sky}}}\big[C_{\ell}^{BB, \rm res}+N_\ell^{BB} + C_{\ell}^{BB, \rm prim}\big]\,,
\label{residual_error}
\end{equation}
where $f_{\rm sky}$ is the sky fraction probed by the observations, which we set to unity in this work; $N^{BB}_\ell$ is the experimental noise; and $C_{\ell}^{BB,\rm prim}$ is the primordial $B$-mode power spectrum, which is negligibly small compared to the other contributions. Once the experimental sensitivity reaches a certain threshold (around $5\,\mu \rm{K}\,\rm{arcmin}$), the measurement variance becomes dominated by the contribution from lensing. Fortunately, this can be remedied by delensing: the higher the delensing efficiency, the lower $C_{\ell}^{BB, \rm{res}}$, and thus the better the precision with which a primordial contribution can be constrained.

\subsection{Impact of non-linear galaxy clustering on delensing}
The galaxy overdensity field can be heuristically described as (see e.g. Ref.~\cite{desjacques_large-scale_2018}):
\begin{align}
\delta_g = & \sum_{\mathcal{O}} \left( b_{\mathcal{O}} + \epsilon_{\mathcal{O}} \right)\mathcal{O}[\delta_m] + \epsilon + \rm{nonlocal} 
\end{align}
where $\mathcal{O}$ denotes operators that non-linearly process the matter density field, and terms proportional to $\epsilon$ denote stochastic contributions which are not correlated with the large-scale matter density field ($\epsilon_{\mathcal{O}}$ introduces stochasticity into the otherwise deterministic $b_{\mathcal{O}}$ bias parameters). In addition, 
there are higher-order derivative terms that account for the fact that galaxy formation is non-local in time through for example the merger history; we leave these implicit for simplicity. We thus see that the galaxy field traces the matter field that caused the CMB lensing distortions --- see equation~\eqref{eqn:kappa_ito_delta} --- but it does so in a non-linear, non-local, and stochastic way.

In this paper, we will study two qualitatively different consequences of this. One is that the non-linearity and stochasticity of galaxy bias can potentially decorrelate fluctuations in the number density of galaxies from the fluctuations in the total matter density which are responsible for the lensing distortions of the CMB. This can make it hard to model the cross-correlation between our tracer and CMB lensing, an essential ingredient of typical models such as equation~\eqref{eqn:full_model}. We will investigate this further in section~\ref{sec:forecasts}.

An additional complication comes from the fact that non-linear gravitational evolution and galaxy bias induce non-Gaussianity in $\kappa^{\rm proxy}$ by correlating the phase of the fluctuations. This can in turn couple with the non-Gaussianities of the CMB lensing potential, which are small but potentially non-negligible, and similarly arise due to large-scale structure non-linearity as well as post-Born lensing~\cite{ref:lewis_challinor_review, fabbianCMBLensingReconstruction}.
This coupling can cause the residual lensing $B$-mode power spectrum to deviate from equation~\eqref{eqn:full_model}, which assumed Gaussianity. These deviations require corrections to the lensing $B$-mode template auto-spectrum and its cross-correlation with the observed $B$-modes (both of which appear implicitly in the power spectrum of delensed $B$-modes) which at leading order arise due to contributions from the lensing convergence bispectrum. Schematically, if the modeling residual is defined as $\Delta C_{\ell}^{BB, \mathrm{res}} \equiv \hat{C}_{\ell}^{BB} - C_{\ell}^{BB, \rm model}$ these corrections have the form
\begin{equation}\label{eqn:bias_nG}
    \Delta C_{\ell}^{BB, \mathrm{res}} = -2\,\Delta C_{\ell}^{\tilde{B} \tilde{B}^{\rm temp}} + \Delta C_{\ell}^{\tilde{B}^{\rm temp} \tilde{B}^{\rm temp}} \,,
\end{equation}
with dominant contributions from the following couplings of the connected trispectrum:\footnote{We have followed the notation of Ref.~\cite{baleato_lizancos_impact_2022}, where $\tilde{X}[Y,\kappa]$ represents the functional dependence of lensed field $\tilde{X}$ on the unlensed field $Y$ and $\kappa$; recall that $\tilde{X}$ is linear in $Y$, so where $\kappa$ is uncontracted, the unlensed field $Y$ is implied. We then use another set of over-bars to denote which unlensed fields are coupled together inside the $n$-point function. On the other hand, bars underneath the expression identify which convergence fields are coupled together into bispectra.}
\begin{equation}
\bcontraction[1.5ex]{\Delta C_{\ell}^{\tilde{B}  \tilde{B}^{\rm temp}} \supset \langle \tilde{B}[E,}{\kappa}
{](\bm{\ell}) \mathcal{B}\big[\tilde{E}[E,}
{\kappa}
\bcontraction[1.5ex]{\Delta C_{\ell}^{\tilde{B}  \tilde{B}^{\rm temp}} \supset \langle \tilde{B}[E,}{\kappa}
{](\bm{\ell}) \, \mathcal{B}\big[\tilde{E}[E, \kappa],}
{\kappa}
\wick[offset=1.5em]{\Delta C_{\ell}^{\tilde{B}  \tilde{B}^{\rm temp}} \supset  \langle \tilde{B}[\c1 E,\kappa](\bm{\ell}) \, \mathcal{B}\big[\tilde{E}[\c1 E, \kappa], \kappa^{\rm proxy}\big](\bm{\ell'})\rangle'_c} 
\end{equation}
and
\begin{widetext}
\begin{equation}
\bcontraction[1.5ex]{\Delta C_{\ell}^{\tilde{B}^{\rm temp}  \tilde{B}^{\rm temp}} \supset 2\, \langle \mathcal{B}\big[\tilde{E}[E,\kappa], }
{\kappa}
{^{\rm proxy}\big](\bm{\ell}) \, \mathcal{B}\big[\tilde{E}[E,}
{\kappa}
\bcontraction[1.5ex]{\Delta C_{\ell}^{\tilde{B}^{\rm temp}  \tilde{B}^{\rm temp}} \supset 2\, \langle \mathcal{B}\big[\tilde{E}[E,\kappa], }
{\kappa}
{^{\rm proxy}\big](\bm{\ell})\, \mathcal{B}\big[\tilde{E}[E,\kappa], }
{\kappa}
\wick[offset=1.5em]{\Delta C_{\ell}^{\tilde{B}^{\rm temp}  \tilde{B}^{\rm temp}} \supset 2\,\langle \mathcal{B}\big[\tilde{E}[\c1 E,\kappa],
\kappa^{\rm proxy}\big](\bm{\ell})\, \mathcal{B}\big[ \tilde{E}[\c1 E,\kappa], \kappa^{\rm proxy}\big] (\bm{\ell}') \rangle'_c}\,.
\end{equation}
\end{widetext}
Note that we have written $\tilde{B}[E,\kappa]$ as a functional analogous to $\mathcal{B}$, but in this case related to the actual lensed $B$-modes. We emphasize that all of these terms vanish if the convergence is Gaussian. 

We quantify the extent to which non-Gaussianity induced by non-linear galaxy bias, structure growth, and post-Born lensing of the CMB give rise to these corrections in section~\ref{sec:nongaussianities}. Similarly, we study the impact of the decorrelation due to non-linear and stochastic bias in section~\ref{sec:forecasts}. We do so using simulated maps obtained from post-processed $N$-body simulations. The equations above therefore serve as a heuristic explanation of the physical origin of the dominant contributions we expect and are not intended as an actual model. 

\section{Simulations and test configuration}
\label{sec:simulations}
We use a fully non-linear cosmological $N$-body simulation to compare the performance of the standard delensing pipeline with a controlled, more realistic case. We use the results from the \textsc{Agora}\footnote{\url{https://yomori.github.io/agora/index.html}} multi-component sky simulations~\cite{omori_agora_2022} for the full-sky lensed $E$-mode and $B$-mode polarization anisotropies, as well as the associated CMB lensing convergence maps and density shells. \textsc{Agora} uses a lightcone built from the  snapshots of the MDPL2 cosmological simulation, which has $(3840)^3$ particles in a volume of $1\, [{\rm Gpc}/h]^3$ for a mass resolution of $1.51\times 10^9\, M_\odot/h$.

The halo catalog is built using the \textsc{Rockstar} phase-space halo finder~\cite{Rockstar}, and astrophysical properties such as stellar masses and star-formation rates are derived from \textsc{UniverseMachine}~\cite{Universemachine}. In particular, we use the non-Gaussian CMB lensing convergence map as well as a realization (`seed 1') of the lensed TQU maps at $\texttt{Nside}=8192$. The simulated maps from \textsc{Agora} have been shown to reproduce CMB lensing observations, as well as many other CMB secondary anisotropies and extragalactic foregrounds~\cite{omori_agora_2022}.

We then construct a coherent realization of the galaxy distribution from the same underlying halo-distribution lightcones using \textsc{SkyLine}\footnote{\href{https://github.com/kokron/skyline}{https://github.com/kokron/skyline}}~\cite{skyline}. We aim to build a full-sky realization of galaxies similar to those that will be targeted by Rubin LSST~\cite{ref:lsst}. This being a magnitude-limited photometric survey, we do not distinguish between quenched and star-forming galaxies, and filter halos in terms of their stellar mass, according to the redshift distribution of the Rubin LSST Gold galaxy sample:
\begin{eqnarray}
   \frac{{\rm d}N}{{\rm d}z}\propto \frac{1}{2z_0}\left(\frac{z}{z_0}\right)^2e^{-z/z_0}\,,
\end{eqnarray}
with $z_0=0.3$ and a mean number density of $\bar{n}=27.3\,{\rm arcmin}^{-2}$.\footnote{This number density is smaller than the nominal expectation for the \emph{gold} sample ($\bar{n}=40\,{\rm arcmin}^{-2}$). The number density of galaxies in our lightcones is limited by mass resolution and assumptions in the astrophysical modeling, since we use the same lightcones as in Ref.~\cite{Kokron:2024ioy}, for which cuts in halo mass and infrared luminosity were applied  to avoid `artificial' satellites without real clustering properties introduced by UniverseMachine to match global and Poisson-like properties. We 
have checked that the increased shot noise (which is larger by a factor of $\sim 1.5$ assuming Poissonian shot noise) does not have a relevant role over the scales of interest and that our conclusions are unchanged. We show further details in appendix~\ref{appendix:shot_noise}.}

The \textsc{Agora} lensed CMB maps we use are produced by ray-tracing through the $N$-body simulation; in this way, they contain contributions from both the non-Gaussianity of large-scale structure and post-Born lensing~\cite{ref:lewis_pratten_16, marozzi_etal_16, marozzi_18}. 
In order to maintain coherence in the perturbative order when going beyond the Born approximation, we remap the galaxy positions based on the convergence calculated out to their position along the line of sight.\footnote{Figure 2 of Ref.~\cite{robertsonDetectableSignalsPostBorn2024} highlights the importance of remapping the galaxy positions when going beyond the Born approximation in order to avoid decorrelation; see also Refs.~\cite{fabbianCMBLensingReconstruction, boehmLensingCorrectionsGalaxylensing2019}. This remapping is sufficiently accurate when done in the Born approximation, so we adopt this approach for simplicity. Note that Ref.~\cite{Namikawa:2016jff} does not remap the galaxy positions in their analysis.} 

In addition, we need to account for photometric redshift errors in our simulated maps. Photometry involves an error $z_{\rm ph}$ in the redshift determination, so that the observed redshift is $z_{\rm obs} = z_{\rm cosmo}+z_{\rm pec}+z_{\rm ph}$, where the first two contributions correspond to the redshift from the expansion of the Universe and the peculiar velocity of the halo, respectively. For each observed galaxy, $z_{\rm ph}$ is expected to be a random draw from a Gaussian distribution centered at zero and with standard deviation $\sigma_z = 0.05(1+z_{\rm cosmo}+z_{\rm pec})$. Finally, we bin our simulated photometric galaxy catalog into 13 different redshift bins according to the values of $z_{\rm obs}$ and edges $z_{\rm edge}=\lbrace 0,\, 1,\, 1.2,\, 1.4,\, 1.6,\, 1.8,\, 2,\, 2.3,\, 2.6,\, 3,\, 3.5,\, 4,\, 4.5,\, 5\rbrace $.\footnote{This redshift binning, which can also be found in e.g., Ref.~\cite{Qu:2022xow}, provides enough flexibility to better match the CMB lensing kernel by independently weighting each galaxy redshift bin.}

However, galaxies need to be deflected  according to the convergence field up to their true radial positions, rather than those corresponding to $z_{\rm obs}$. This would not be a problem if we applied the deflection at the catalog level, i.e., on a halo-per-halo basis. However, in order to minimize the computational expense, we apply the deflection of the galaxy positions at the map level using \texttt{lenspyx}\footnote{\href{https://github.com/carronj/lenspyx/tree/master}{https://github.com/carronj/lenspyx/tree/master}}~\cite{Reinecke:2023gtp}. For each source redshift, we must therefore calculate a different deflection/convergence field from the density shells provided with the \textsc{Agora} suite (we do this following section III.B of~\cite{robertsonDetectableSignalsPostBorn2024}), and apply it as we generate the number-count maps for each of the redshift bins listed above.

In practice, we generate our simulated maps as follows. As we iterate over all the files $f_j$ containing the halos in the lightcone, where $j$ indexes the redshift shell, we:
\begin{enumerate}
    \item load the halos in $f_j$ and apply the photometric redshift error as specified above;
    \item assign the halos to different empty $M_{\rm gal}^{(i,j)}$ \texttt{Healpix} maps with $N_{\rm side}=8192$ according to $z_{\rm obs}$, where $i$ indexes the redshift bin between $z_{\rm edge}$ indicated above;
    \item load the convergence map from \textsc{Agora} (which we have downgraded to $N_{\rm SIDE}=2048$) corresponding to $f_j$ as the source redshift and apply the deflection to each $M_{\rm gal}^{(i,j)}$
    \item Add $M_{\rm gal}^{(i,j)}$ to the final redshift-bin galaxy number counts maps $M_{\rm gal}^i$.
\end{enumerate}

All these points can be schematically summarized with the following expression for the number-count maps $M^i_{\rm gal}$:
\begin{equation}
 M^i_{\rm gal}(\hat{\pmb{n}})=\sum_j\kappa_j(\hat{\pmb{n}}')\diamond M_{\rm gal}^{(i,j)}(\hat{\pmb{n}}')\,,
\end{equation}
where $\hat{\pmb{n}}$ and $\hat{\pmb{n}}'$ are the deflected and original positions, and $\diamond$ denotes the lensing deflection of a scalar map $M$ due to a convergence field $\kappa$, computed using \texttt{lenspyx}. 

\begin{figure}[hbt!]    \includegraphics[width=\columnwidth]{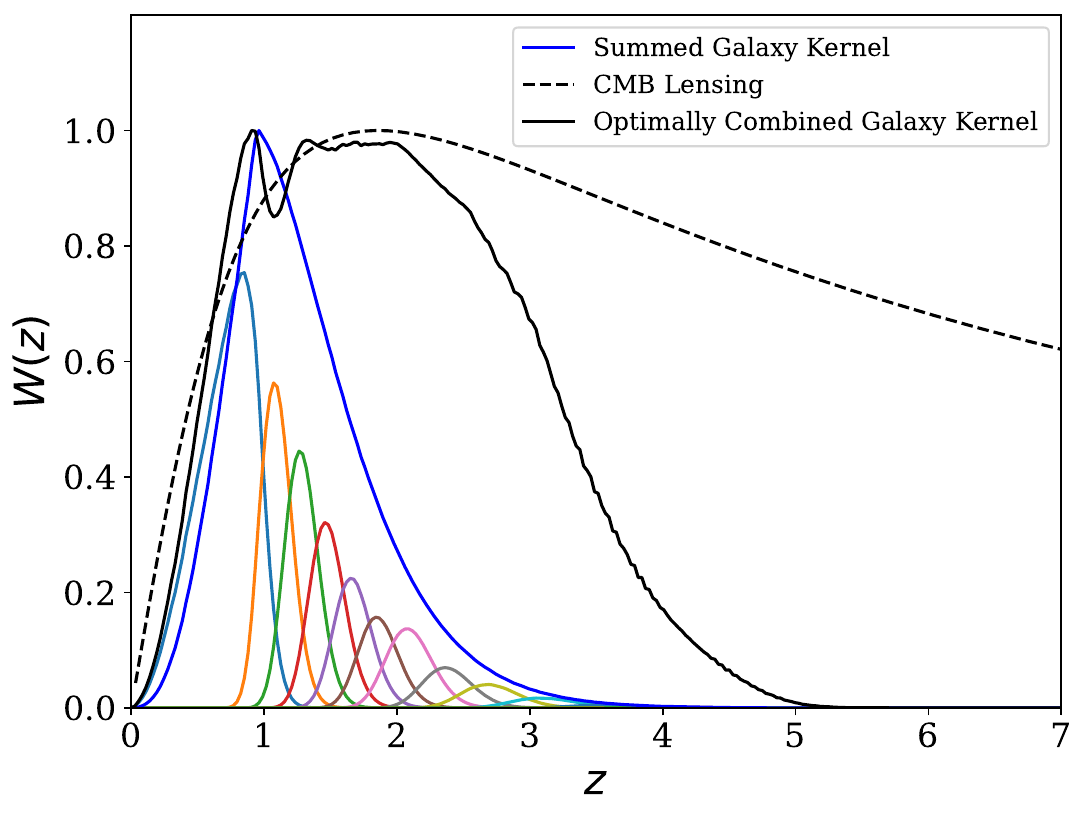}
    \caption{\label{fig:kernels} Projection kernels for the galaxy overdensity (solid) and CMB lensing (dashed, rescaled to match the amplitude of the galaxy kernels). The solid, dark blue curve shows the lump-sum of all the galaxies into a single bin, while the solid black curve shows the kernel for the optimal combination of bins at $\ell=500$, computed following equation~\eqref{combine_weights}.}
\end{figure}

The \textit{true} redshift (i.e., $z_{\rm cosmo}+z_{\rm pec}$) distributions of the resulting samples, multiplied by a value of linear bias that we fit for (see section~\ref{sec:forecasts}) is shown in figure~\ref{fig:kernels}. We also show the same quantity for the combination of all galaxies into a single redshift bin as well as the kernel for the optimal combination on bins from equation~\eqref{combine_weights} for modes with $\ell=500$. For reference, we show also the CMB lensing kernel, normalized to the same amplitude.

In all cases we compute angular power spectra of any combination of maps using the publicly-available package \texttt{healpy}\footnote{\href{https://healpy.readthedocs.io/en/latest/}{{https://healpy.readthedocs.io/en/latest/}}}. Note that, since the underlying density field of the lightcones employed to generate the simulated maps we use in this work are taken from cubic snapshots of $1\, {\rm Gpc}/h$ side, there are no correlations at scales $\gtrsim 1\, {\rm Gpc}/h$. Therefore, we cannot study the largest angular scales. This is no limitation for our study, since the lenses that source the large-scale lensing $B$-modes subtend angular scales $200\lesssim \ell \lesssim 800$ (see e.g., Ref.~\cite{ref:smith_12_external}), which at the redshifts of these galaxy samples correspond to spatial scales smaller than 1\,Gpc/$h$.

\section{Results}\label{sec:results}
\begin{figure}[hbt!]    \includegraphics[width=\columnwidth]{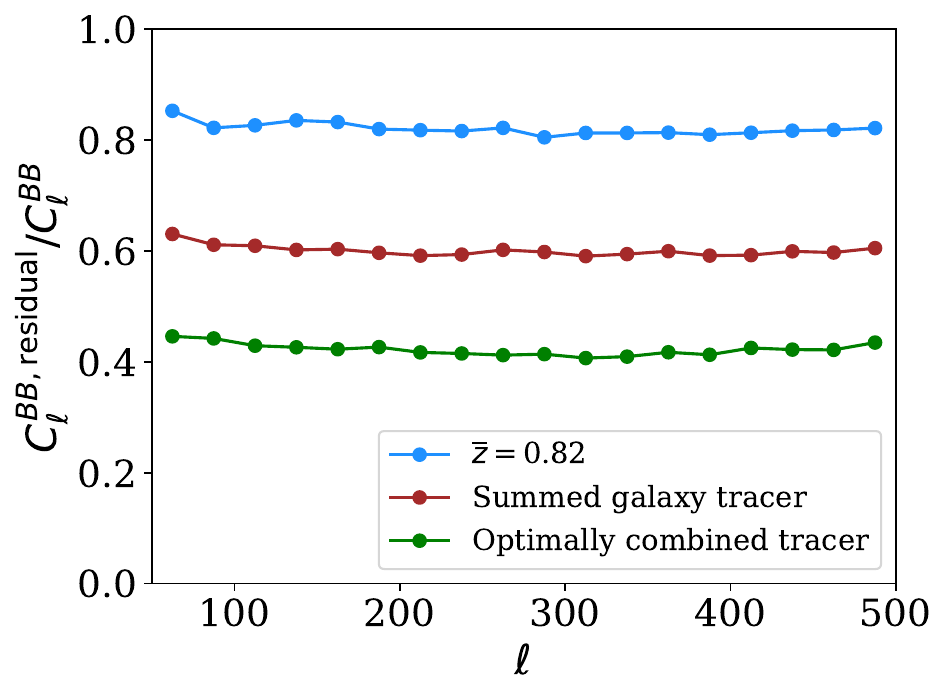}\caption{\label{fig:residuals_combined}Fractional residual lensing $B$-mode power spectrum after delensing using our realistic, simulated galaxies. Different curves correspond to different combinations of galaxies with the redshift distributions shown in figure~\ref{fig:kernels}: galaxies from the lowest redshift bin (blue), the brute-force combination of all galaxies in every redshift bin into a single one (brown), or the optimal weighting of all the individual redshift bins (green).}
\end{figure}

Equipped with the simulations and the delensing infrastructure described in previous sections, we are now ready to assess delensing performance. To derive some intuition, we show in figure~\ref{fig:residuals_combined} the residual lensing $B$-mode power after applying map-level delensing using different combinations of our simulated galaxy samples; see appendix~\ref{appendix_curved} for details about our curved-sky implementation. 

Naturally, different redshift bins of the galaxy sample correlate to different extents with the CMB lensing potential and thus remove different amounts of lensing $B$-modes when used as part of delensing pipelines. It is instructive to consider the case of the bin centered at $\bar{z}=0.82$, which is the lowest-redshift and densest of the ones we consider. We see that by itself it is able to remove approximately $20\%$ of the large-scale lensing $B$-mode power. Higher-$z$ bins are in principle better tracers of the CMB lensing kernel, as it peaks around $z\sim2$. However, galaxies in them are also typically sparser, and therefore less signal-dominated, on the scales that matter when delensing large-scale $B$-modes ($200\lesssim L \lesssim 800$; see e.g.~\cite{ref:smith_12_external}). So it is not surprising that including higher-redshift bins improves delensing performance as seen in the figure. Note, however, that using the optimal weights of equation~\eqref{combine_weights} (green curve) leads to better delensing efficiency than lumping all the galaxies together into a single bin (brown curve in the figure). The reason is that the optimal combination exploits the freedom to upweight whichever bins are more correlated with CMB lensing on a given angular scale.

These trends are all well studied and characterized~\cite{ref:smith_12_external, ref:sherwin_15, ref:yu_17, ref:cib_delensing_biases, ref:manzotti_18, ref:SO_delensing_paper}. Let us now focus instead on something less understood: the impact of non-linearities associated with galaxy bias and gravitational collapse. Throughout our investigation, we will explicitly consider the two binning scenarios above, where we either lump all the galaxies into a single set or we weight different bins optimally before combining them together.

\subsection{Model accuracy assuming \\ Gaussian galaxy distributions}\label{sec:nongaussianities}
A key ingredient of any delensing pipeline is a model for the residual lensing $B$-mode power spectrum. In a $\Lambda$CDM cosmology, the lensing $B$-mode power spectrum can be predicted very accurately. Thus, the spectrum after delensing can be predicted equally accurately if $\kappa^{\rm proxy}$ and its cross-correlation with CMB lensing are well characterized. Typical models for the residual lensing $B$-mode power spectrum assume $\kappa^{\rm proxy}$ is Gaussian-distributed and take the form of equation~\eqref{eqn:full_model}. In practice, rather than using that simple analytic expression, the model is obtained by applying the pipeline to Gaussian simulations of the large-scale structure tracer to more easily incorporate observational effects (see, e.g., Refs.~\cite{ref:SO_delensing_paper, hertig_simons_2024}). Either way, non-Gaussianities arising from non-linear structure formation and galaxy bias
can generate additional contributions of the form of equation~\eqref{eqn:bias_nG} not captured by these models. Note that the non-Gaussian features sourced by nonlinear matter-density evolution were already studied in Ref.~\cite{ref:namikawa_16}.\footnote{Though, as noted above, the perturbative consistency may have been compromised by working with ray-traced lensed CMB maps and un-deflected galaxy positions.} Here, we consider biased tracers, thus allowing for an additional source of non-Gaussianity.

To quantify the size of these corrections, we now compare the residual lensing $B$-mode power after carrying out map-based delensing --- following appendix~\ref{appendix_curved} --- using two variations of the mass proxy. On the one hand, we use the galaxy mocks described in the previous section, which are built around $N$-body simulations and thus inherit a range of effects from non-linear clustering down to galaxy stochasticity. We will refer to this case as the `non-Gaussian simulation' or our `ground truth'. 

\begin{figure}[hbt!]    \includegraphics[width=\columnwidth]{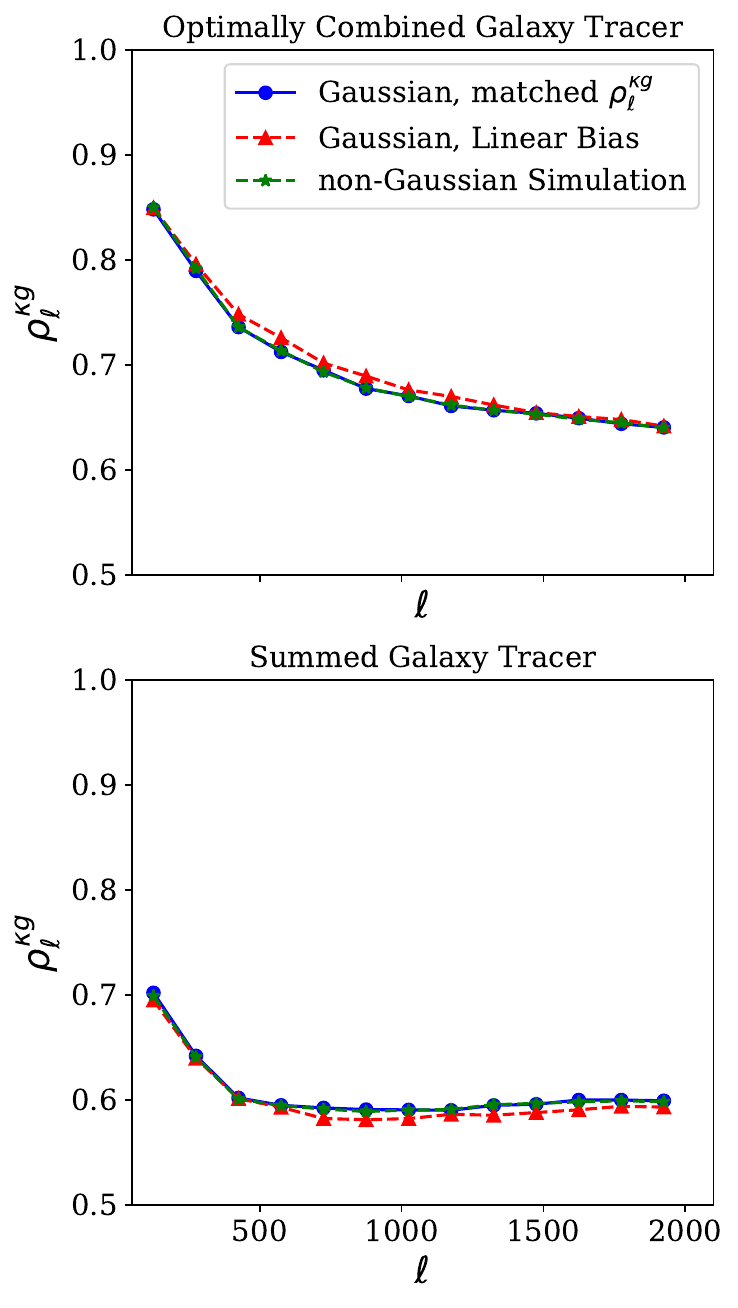} \caption{\label{fig:rho} Correlation coefficient with CMB lensing of various samples of simulated galaxies. In the top panel, different tomographic bins are combined optimally; in the bottom one, they are simply added together into a single map. The green, dashed curves show our 'ground truth', measured from the non-Gaussian simulation. By construction, this overlaps with the solid, blue curve, though the latter pertains to a Gaussian simulation. The red, dashed curve also corresponds to a Gaussian simulation, albeit one that further assumes linear galaxy bias with a value fit to the large-scale clustering of the non-Gaussian mock. Though the latter is clearly not a good approximation for $\ell$ larger than a few hundred, these scales are relatively uninmportant when delensing low-$\ell$ $B$-modes.}
\end{figure}

On the other, we use Gaussian simulations (produced following appendix F of Ref.~\cite{ref:cib_delensing_biases}) which by construction have the same average angular auto-spectra, cross-spectra with CMB lensing, and cross-spectra between galaxy redshift bins as the mocks produced directly from the $N$-body simulations. These Gaussian simulations are built on top of a Gaussian CMB lensing convergence provided as part of the \textsc{Agora} suite; naturally, we use the lensed CMB fields obtained from remapping using this lensing map. The two cross-correlation coefficients are shown in figure~\ref{fig:rho} --- they are indeed indistinguishable.

\begin{figure}[hbt!]    \includegraphics[width=\columnwidth]{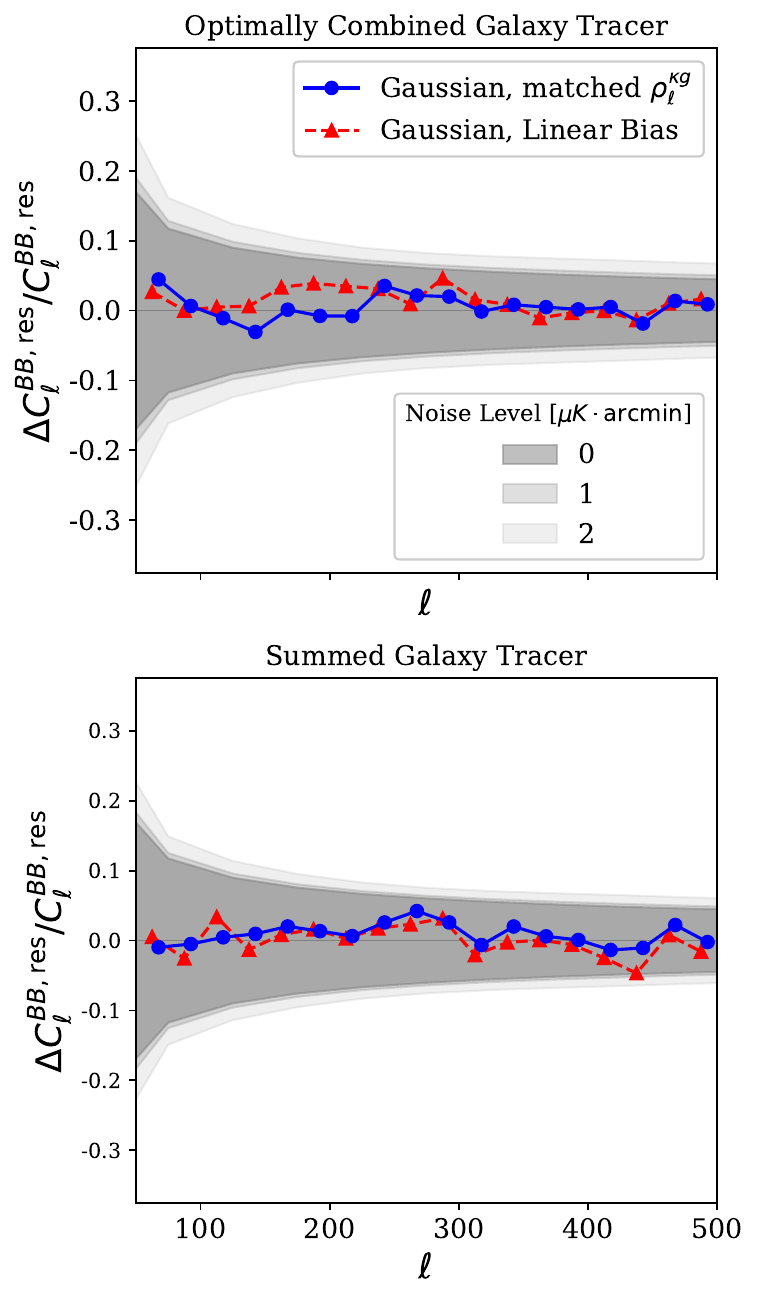}\caption{\label{fig:residual_plot}Fractional difference between the residual lensing $B$-mode power spectrum obtained after delensing with our realistic non-Gaussian simulations and either of two approximate treatments: one involving a Gaussian simulation which by construction has the same correlation with CMB lensing as our `ground truth' simulation (blue), and another where the Gaussian simulation involves the additional assumption that galaxy bias is linear and is measured from the large scale clustering of the non-Gaussian mocks (red). In the top panel, different bins of our tomographic sample are combined optimally; in the bottom one, they are all lumped together into a single bin. The shaded regions denote the $1\,\sigma$ fractional (Gaussian) uncertainty assuming full-sky coverage and various white noise levels for the large-angle polarization measurements.}
\end{figure}

Since we are enforcing by construction that the two sets of mocks have the same correlation coefficient with CMB lensing and the delensing weights are also equal by construction, the delensing performance arising from their Gaussian part --- given by equation~\eqref{residual_power_spectrum_flat} --- is expected to match, while any differences can be attributed to non-Gaussian couplings. We investigate this in figure~\ref{fig:residual_plot}, where we compare the performance of both delensing pipelines at the level of the residual lensing $B$-mode power spectrum. Both when we lump all galaxies into a single set, and when we weight different bins optimally before combining them together, we find that any deviations in the $B$-mode power spectrum post-delensing are smaller than cosmic variance. Perhaps more importantly, the residual is consistent with noise fluctuations around zero, showing no clear shape that could be confused for a primordial contribution to the $B$-mode power spectrum. 

Our findings are consistent with Ref.~\cite{ref:namikawa_and_takahashi_19}, who characterized the corrections to delensing associated with non-Gaussianities coming from the non-linear growth of structure (assuming linear galaxy bias) and found them to be below percent-level. This is also the approximate size of the contribution to the lensing $B$-mode spectrum from non-Gaussianities associated with non-linear large-scale structure and post-Born corrections~\cite{ref:lewis_pratten_16}.

The relatively small deviation is likely due to projection along the radial direction. Even though the 3D fluctuations that source the large-scale lensing $B$-modes are in principle affected by non-linear clustering and bias ($k\lesssim0.4$\,Mpc\,$h^{-1}$ at $z\sim1$), the angular galaxy-number-density fluctuations are Gaussianized due to the projection along the line of sight over the wide redshift bins afforded by photometric surveys.

These results indicate that in any inference pipeline aiming to detect a primordial $B$-mode contribution, the lensing residual can be modeled sufficiently well as long as the cross-correlation coefficient between CMB lensing and the mass tracers is determined accurately (possibly from the measured auto-and cross- spectra of the tracers with CMB lensing reconstructions).

\subsection{Model accuracy assuming \\ linear galaxy bias}\label{sec:forecasts}
We have established in the previous subsection that models for the lensing $B$-mode residual power spectrum that assume Gaussianity of the lensing convergence proxy are actually highly accurate. This means that, as long as $C_{\ell}^{\kappa \kappa}$, $C_{\ell}^{\kappa \kappa_{\rm proxy}}$ and $C_{\ell}^{\kappa_{\rm proxy} \kappa_{\rm proxy}}$ are close to the truth, equation~\eqref{eqn:full_model}
returns an accurate prediction of the power spectrum of residual lensing $B$-modes after delensing\footnote{Note that evaluating equation~\eqref{eqn:full_model} requires knowing the spectra individually. On the other hand, if the delensing weights are close to the truth such that equation~\eqref{residual_power_spectrum_flat} holds, then only their combination into the corss-correlation coefficient ($\rho$) matters. (Knowledge of the individual spectra fixes the correlation coefficient, $\rho$, but the converse is not true.)}. 

In practical applications, these spectra will need to be determined empirically from fits to noisy measurements of the tracers' auto- and cross-correlations with internal reconstructions of the CMB lensing convergence (see Ref.~\cite{ref:SO_delensing_paper} for an extensive discussion). Though the exact fitting form used does not matter --- the best-fit values of the model parameters will not be interpreted in any way and only an accurate estimation of the cross-correlation %
matters --- it is important to use as simple a model as possible to avoid overfitting measurement noise and systematics.

In order to capture the difficulty of accurately characterizing the tracer spectra from data, we investigate the case where the angular spectra of galaxies are modeled assuming linear galaxy bias. This is intended to reflect a maximally conservative scenario where systematics and/or measurement noise (primarily in the internal CMB lensing reconstruction needed to determine $C_{\ell}^{\kappa \kappa_{\rm proxy}}$)\footnote{For SO, for example, the signal-to-noise ratio per mode on the lensing auto-spectrum drops below unity for $L\gtrsim250$~\cite{ref:SO_science_paper} --- the variance of the cross-spectrum is linearly proportional to this. The auto-spectrum of $\kappa^{\rm proxy}$ is expected to have higher signal-to-noise, but systematics may also present challenges; see Ref.~\cite{ref:SO_delensing_paper} for an extensive discussion).} make it such that the models can only be fit to a limited range of large-scale multipoles. The fit may then only be sensitive to the linear galaxy bias, leaving the impact of non-linear bias and stochasticity unmodeled and thus compromising the extrapolation to smaller scales. In this section, we set out to gauge the impact of this last point. We refer the reader to Ref.~\cite{ref:SO_delensing_paper} for a study of how measurement uncertainty affects the determination of tracer spectra (assuming linear galaxy bias) and thus delensing.

This exercise has implications for how well the post-delensing $B$-mode power spectrum can be modeled, but also for the accuracy of forecasts of delensing performance, which traditionally use the model in equation~\eqref{residual_power_spectrum_flat} and assume linear galaxy bias\footnote{Equation~\eqref{residual_power_spectrum_flat} assumes that the fiducials going into the delensing weights match the truth. Though this is not strictly the case for the `ground-truth' side of the comparison in this section, we verify thar the deviation is very small, and at any rate this is a second order effect on delensing performance.} (see e.g., Ref.~\cite{Bellomo:2020pnw}).

In order to quantify these modeling inaccuracies, we compare the residual lensing $B$-mode power after map-level delensing in two scenarios which share the same delensing weights (these are always known by construction) but differ in the underlying correlation between the galaxies and CMB lensing due to the scale-dependent and stochastic nature of galaxy bias. The galaxy mocks derived from the $N$-body simulation will serve as the `ground truth', and we will compare this Gaussian simulations (the accuracy of which in modeling the lensing residuals we have already established) derived from spectra that are fit to the realistic mocks on large scales, but which are extrapolated to small scales using the assumption of linear bias. We will also make the reasonable assumption that cosmology is known by other means and match the non-linear matter power spectrum between the two. This way, we isolate the effect of nonlinear bias, which is our target of study in this test.

In more detail, for each redshift bin of the simulations, we fit the measured angular cross-spectrum between the galaxy overdensity and the noiseless CMB lensing covergence, $C_{\ell}^{\kappa g}$, on large scales $50 < \ell < 150$ (assuming that these scales are largely unaffected by non-linearities in the bias) with the analytical model in equation~(\ref{flat_spectrum}) (as implemented in the Limber approximation in \texttt{PyCCL}),  
using
\begin{eqnarray}
b_{\rm linear}(z) = b_{0} + b_{1}z,
\label{scale-dependant_bias}
\end{eqnarray}
as a parametrization of the linear galaxy bias, where $b_{0}$ and $b_{1}$ are free parameters to be fit to the measured power spectra in the indicated range of scales. Here, $b_1$ accounts for bias evolution with redshift within the bin. We fix the non-linear matter power spectrum to the \texttt{HaloFit} prediction in the MDPL2 cosmology,\footnote{The \texttt{HaloFit} prediction for the non-linear matter power spectrum is calibrated to $N$-body simulations, and it is expected to match very well the non-linear matter clustering in the MDPL2 simulation over the scales we are concerned with here. } %
and use the redshift distributions described in section~\ref{sec:simulations}. We then use this value of galaxy bias, together with the redshift distribution and an analytically-determined Poissonian shot noise component\footnote{Note that if the shot noise is in reality not Poissonian, this may source further discrepancy between the two pipelines.} to generate galaxy density auto power spectra following equations (\ref{flat_spectrum}) and (\ref{shot_noise}). From the linear bias of each bin, we also obtain a prediction for the cross-spectrum of all bin pairs with no new free parameters. 

Since we already have verified the validity of the Gaussian approximation for $B$-mode delensing in the previous section, now we use the $C_{\ell}^{g_{i}g_{j}}$ and $C_{\ell}^{\kappa g_{i}}$ computed with the assumptions described above to generate Gaussian realizations with the appropriate correlations.\footnote{The Gaussian galaxy mocks we produce are built on top of the Gaussian CMB lensing potential provided within the \textsc{Agora} suite.}
This `approximate' galaxy sample  constitutes a proxy for our modeling attempts in real analyses, to be compared with our `ground truth' simulations, which are in turn a proxy for the actual data. We also use these `approximate' spectra to calculate the optimal weights $c_i$ in equation~\eqref{combine_weights}. In order to disentangle our findings from a simple difference in delensing efficiency due to the choice of weights, we apply these weights to both the Gaussian and non-Gaussian/ground-truth mocks.

To get some insight, we can calculate $\rho_\ell^{ab}$ from these spectra and compare it to the `ground truth'; this is shown in figure~\ref{fig:rho}. As expected, the linear bias assumption (red curves) misestimates the correlation coefficient away from large angular scales. But note that when it comes to delensing large-scale $B$-modes, it is only lenses with $200\lesssim L \lesssim 800$ that matter.

This latter point is in fact made more directly by figure~\ref{fig:residual_plot}, where the red lines show the fractional difference in the residual lensing $B$-mode power spectrum after delensing with the `ground truth' simulation or with our Gaussian simulation with the linear bias assumption baked in (recall that $\Delta C_{\ell}^{BB, \rm res}$ is positive when the linear bias model underpredicts the measured $B$-mode power). Let us focus on the `optimally-combined' scenario for concreteness: as a consequence of the correlation with CMB lensing being overestimated when working under the linear bias assumption, this model slightly over-predicts the delensing efficiency on the relevant scales, leaving a positive $\Delta C_{\ell}^{BB, \rm res}$.

It appears that on the relevant scales the impact is marginally more acute when galaxy bins are optimally combined together. This may be because higher-redshift bins closer to the peak of the CMB lensing kernel get upweighted with respect to the case in which all galaxies are simply added. Though this enables more extensive delensing (cf. figure~\ref{fig:residuals_combined}) it also brings to the fore the shortcomings of the bias approximation, as all other things being equal, galaxies are more biased tracers of the matter at higher redshift.

For our Rubin-like galaxies, the optimal combination results in a modelling residual that is at most $\sim 5\%$ of the lensing $B$-mode power spectrum amplitude after delensing, smaller than the $1\sigma$ uncertainty associated with cosmic variance on a per-mode basis. Nonetheless, we observe that this deviation with respect to our `ground truth' is systematic, so that it may have an impact in the inferred value of $r$ despite lying below cosmic variance on a per-mode basis. Note that, since we are considering a full-sky scenario, the only way to improve over the cosmic variance considered here is to delens more extensively, be it using internal reconstructions or external tracers that better correlate with the CMB lensing potential. 

\subsection{Bias in the inference of $r$}

It is worth noting that in addition to being small, the fractional residual in our plots is relatively flat. This is good news from the perspective of inference, as a residual that has the same shape as the lensing $B$-mode can easily be absorbed into a lensing amplitude parameter during marginalization, and it is unlikely to be mistaken for a primordial component, which has a markedly different shape (any goodness of fit statistic should return a poor result). 

The physical origin of such a flat residual likely has to do with the fact that the deviations in $\rho^{\kappa g}$ appear chiefly on small scales while the large scales are `correct' by construction. We are therefore mostly mis-estimating the extent to which small-scale lenses are removed during delensing. It pays to understand the role that these small-scales lenses play: since they are not correlated over large angles, their impact on the large-scale polarization anisotropies has to look like scale-independent white noise. Any small-scale mis-modelling will thus only manifest on large scales as a correction to the residual lensing $B$-mode amplitude. Nevertheless, this amplitude mismatch may impact the significance of any potential detection of primordial tensor-to-scalar ratio $r$, or even bias its measurement.

In order to quantify the impact of our various approximations on measurements of the tensor-to-scalar ratio, $r$, we estimate the difference $\Delta\hat{r}\equiv r_{\mathrm{inferred}} - r_{\mathrm{true}} $ between its \textit{inferred} and \textit{true} value. It can be shown using the Fisher-matrix approximation (see e.g.~\cite{Roy_Kulkarni_Meerburg_Challinor_Baccigalupi_Lapi_Haehnelt_2021, Bernal:2020pwq}) that the maximum-likelihood estimate for the parameter $r$ is biased by the following amount:
\begin{equation}
    \Delta\hat{r}=\frac{\sum_{\ell_{\rm min}}^{\ell_{\rm max}}\Delta C_{\ell}^{BB} \ C_{\ell}^{BB,\rm prim}(r=1)/\sigma_{\rm res}^2(\ell)}{\sum_{\ell_{\rm min}}^{\ell_{\rm max}}[C_{\ell}^{BB,\rm prim}(r=1)]^{2}/\sigma_{\rm res}^2(\ell)}\,,
    \label{eq:delta_r}
\end{equation}
where $\Delta C_{\ell}^{BB}\equiv \hat{C}_{\ell}^{BB} - C_{\ell}^{BB, \mathrm{model}}$ is the unmodeled residual. In our case, $\Delta C_{\ell}^{BB} = \Delta C_{\ell}^{BB,\rm res} =  C_{\ell}^{BB,\rm ground \, truth} -  C_{\ell}^{BB,\rm approximation} $ is the difference between the residual lensing $B$-mode power spectrum obtained after delensing with the mock galaxy sample drawn from the $N$-body simulation (our `ground truth') and the spectrum obtained after delensing under the various approximations we have considered (see figure~\ref{fig:residual_plot}). For the per-mode power spectrum variance, $\sigma^2_{\rm res}$, we use the Gaussian approximation, including white noise and a non-zero primordial signal (indicated explicitly where relevant). We evaluate this expression between the limiting multipoles $\ell_{\rm min}=30$ and $\ell_{\rm max}=300$; there is little signal at higher multipoles, and lower ones are difficult to access due to systematics (particularly for ground-based experiments).

In this work we considered two sources of $\Delta C_{\ell}^{BB,\rm res}$: corrections stemming from the non-Gaussianity of $\kappa$ and the large-scale structure tracers used for external delensing, and a mismodeling of the tracers' auto- and cross-correlation with the CMB lensing potential in light of measurement error and galaxy bias.
From figure~\ref{fig:residual_plot} we can see that the latter effect is likely to be more significant. 

\begin{figure}[hbt!]    \includegraphics[width=0.95\columnwidth]{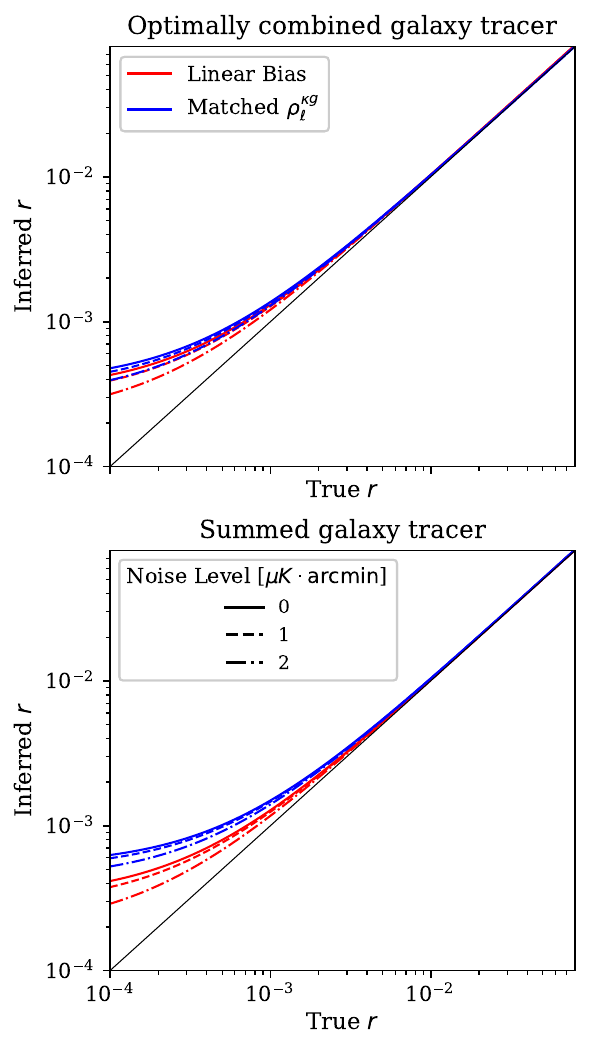}
    \caption{\label{fig:inferred_r} Comparison of the inferred and true values of the tensor-to-scalar ratio, $r$, in the presence of two sources of modeling error: blue curves assume the galaxy distribution is Gaussian, while red curves further assume galaxy bias is linear. In each case, the inference is done in presence of a white noise component with amplitude given in the legend. In the top panel, different redshift bins are optimally combined together using the scale-depedent weights of equation~\eqref{combine_weights}; in the lower panel, they are simply lumped together into a single bin.}
\end{figure}

We show the resulting bias due to these approximations in figure~\ref{fig:inferred_r} for three different levels of white noise: 0, 1 and 2 $\mu\rm{K}\,$arcmin. Note that we are ignoring very important sources of noise on large scales such as atmospheric noise and Galactic foregrounds.\footnote{Exploring the large space of foreground models and experimental configurations is beyond the scope of this work.} Our results can therefore be regarded as an upper limit on the bias on $r$. 

We find that $r$ would be overestimated for both sets of approximations --- i.e., when non-Gaussianities are ignored and when, in addition, linear galaxy bias is assumed. 
However, the bias is small compared to the statistical precision of most upcoming experiments that would use the Rubin-like galaxies we have simulated to delens $B$-modes.  We can compare with the expected $\sigma(r=0)$ achieved by SO after delensing in two different foreground scenarios, as reported by Ref.~\cite{hertig_simons_2024}: in the lowest foreground complexity scenario, $\sigma(r=0)\approx 1.2\times 10^{-3}$, while in the higher-complexity scenario $\sigma(r=0)\approx 2.3\times 10^{-3}$. In both cases, the biases on $r$ are smaller than the 1-$\sigma$ uncertainty.\footnote{Note that the SO forecasts assume more extensive delensing than what is afforded by our simulated galaxies, with the residual lensing $B$-mode spectrum down to $35\%$ of its original value (cf. $\sim 42\%$ in our case according to figure~\ref{fig:residual_plot}). This means that the bias would be even less statistically significant if the delensing performance of SO were matched to what we are achieving with our galaxies.} For more ambitious experiments such as CMB-S4, care will be required to marginalize over these modeling residuals in order to avoid bias in the inference of $r$.

\section{Conclusions}\label{sec:conclusions}
Delensing is already a key ingredient of searches for the imprint of primordial gravitational waves on the $B$-mode polarization of the CMB. Galaxy surveys have been identified as a powerful proxy of the density fluctuations responsible for gravitational lensing of $E$-modes into $B$-modes, and thus hold great promise for delensing. 

In this work, we addressed a key, previously unanswered question that had raised doubts about the ultimate utility of external tracers of the large-scale structure as effective tracers of the matter distribution for $B$-mode delensing: the impact of non-linear clustering and bias. Such non-linearities can in principle induce non-Gaussinity and also decorrelate the galaxy field from the lensing potential, complicating the modeling of the residual $B$-modes after delensing and compromising constraints on primordial gravitational waves if they are significant and go unaccounted for.

We investigated the size of these effects in a mock delensing analysis using a magnitude-limited sample of galaxies resembling the \emph{gold} selection of Rubin LSST. Crucially, we obtained our sample by populating halos in a lightcone from an $N$-body simulation. The sample thus obtained is non-linearly biased relative to the matter and is therefore significantly more realistic than what has been considered previously in the literature. The matter distribution underlying our galaxy samples is itself non-linear and non-Gaussian, and is the same one used to produce the \textsc{Agora} mocks~\cite{omori_agora_2022}. By working with CMB maps that have been lensed by this non-Gaussian lensing potential via ray-tracing, and also accounting for gravitational lensing in calculating the apparent positions of galaxies, we ensure that the impact of non-Gaussianity and non-linearity are accounted for. We then used these realistic galaxy mocks as the `ground truth' against which to compare various modeling approximations that are usually made in delensing analyses.

The first approximation we investigated is that of treating the external biased tracer as a realization of a Gaussian random field; see section~\ref{sec:nongaussianities}. By comparing the outcome of delensing using our ground truth mocks versus using a mock tracer that has on average the same two-point statistics as the `realistic' galaxies but is fully Gaussian, we verified that the standard assumption of Gaussianity leads to no significant changes to the power spectrum of residual lensing $B$-modes after delensing, and negligible bias on $r$ for any foreseeable delensing analysis using the Rubin \emph{gold} sample.

Then, in section~\ref{sec:forecasts}, we investigated a second potential issue related to the fact that galaxy bias decorrelates the tracer from the matter distribution that caused the lensing deflections. From equation~\eqref{eqn:full_model} --- a typical model for the power spectrum of delensed $B$-modes --- it is clear that the galaxies' auto- and cross-correlation with CMB lensing need to be characterized accurately in order to avoid biasing the inference on $r$. However, measurement uncertainty and systematics may compromise our ability to determine these empirically. Ref.~\cite{ref:SO_delensing_paper} presented an extensive investigation of the impact of measurement uncertainty on this characterization. Here, we complement their conclusions by examining a conservative scenario where we assume that only the linear part of the galaxy bias can be determined from the data, and the tracer spectra used in the model have to be computed by extrapolating to smaller scales using the assumption of linear bias. Practically, such modeling inaccuracy can be studied by comparing our ground truth delensing pipeline to another one which employs the same delensing weights but uses as the tracer a Gaussian random field\footnote{Note that we have already established the accuracy of the Gaussian approximation.} with angular power spectra fit to the large scales of the mocks while assuming a model where galaxy bias is linear (and a non-linear matter power spectrum calculated from \texttt{HaloFit} in the same cosmology as the ground truth mocks).

We find that the linear bias assumption misestimates the degree of correlation between the galaxies and the CMB lensing potential on small scales. However, the deviation is small on the scales that are of relevance to delens large-scale $B$-modes. We estimate that this effect leads to a small overestimation of the amplitude of $r$, of order a few times $10^{-4}$ for $r=0$. This is significantly smaller than the 1-$\sigma$ statistical uncertainty expected of CMB experiments that could potentially use these galaxies for delensing --- at most half of the statistical uncertainty expected of SO post-delensing~\cite{namikawaSimonsObservatoryConstraining2022b}. It is possible that the effect could become significant for a more ambitious experiment like CMB-S4. 

We note that these findings also have implications for the accuracy of forecasts of delensing performance, which traditionally use the model in equation~\eqref{residual_power_spectrum_flat} and assume linear galaxy bias. Our results suggest that the decorrelation induced by non-linear and stochastic galaxy bias is relatively small on the relevant scales such that previous forecasts based on the Rubin \emph{gold} sample --- e.g., Refs.~\cite{ref:manzotti_18, ref:yu_17, ref:SO_delensing_paper, hertig_simons_2024, ref:cib_delensing_biases, baleato_lizancos_impact_2022} --- should correctly predict the residual lensing $B$-mode power after delensing with an accuracy better than $5\%$ on a per-mode basis.

The reader may have noted that we have been careful to restrict the applicability of our conclusions to the Rubin \emph{gold} sample. The reason is that the impact of galaxy bias non-linearity and stochasticity will be different for different sample selections. The suitability of different selections for delensing has in fact not been studied to date. The \emph{gold} sample considered here is magnitude-limited, so it contains a large number of intrinsically-dim, low-redshfit galaxies which have relatively low bias, restricting the impact of bias non-linearities. On the other hand, the dynamics of these intrinsically-dim galaxies are often influenced primarily by their local environment, rather than the large-scale structure. An alternative would be to select on color to isolate, for example, a population of `Luminous Red Galaxies' (LRGs) that more faithfully trace the large-scale matter field. However, these objects are highly biased, so their usefulness for delensing may well merit a dedicated study. The ideal selection might be one that comprises relatively low-bias objects with a number density that remains constant with redshift and contains enough galaxies that shot noise remains subdominant down to the smallest angular scales needed for delensing ($L\lesssim 800$). 

In spite of this caveat, our conclusions imply that the impact of non-linearities in matter clustering and galaxy bias on the $B$-mode power spectrum after delensing will be small for any `reasonable' selection of galaxies. More work is however needed to characterize how the non-linearities studied in this paper affect the covariance of the $B$-mode power spectrum after delensing. The pioneering work of Ref.~\cite{ref:namikawa_and_takahashi_19}, who studied this exact question in the case of linear galaxy bias, suggests that the impact is likely to be very small. Further work is also needed to extend these investigations to applications of delensing in searches for cosmic birefringence~\cite{namikawaTomographicConstraintAnisotropic2024}, primordial non-Gaussianity~\cite{coultonParityoddIntrinsicBispectrum2021}, and the delensing of other CMB angular power spectra~\cite{hotinliBenefitsCMBDelensing2021}.

All in all, the results of our work are very encouraging for the prospects of delensing CMB $B$-modes with galaxy surveys. This is good news for upcoming experiments such as the Simons Observatory~\cite{ref:SO_science_paper} or CMB-S4~\cite{ref:S4_forecast}, which will be able to complement their internal lensing reconstructions with these external tracers.

\begin{acknowledgments}
The authors wish to thank the lensing working group of SO --- particularly Blake Sherwin --- for conversations that inspired this project, as well as Giulio Fabbian and Martin White for discussion about non-Gaussian lensing signals. SW thanks the Network for Neutrinos, Nuclear Astrophysics, and Symmetries (N3AS) for support. JLB acknowledges funding from the project UC-LIME (PID2022-140670NA-I00), financed by MCIN/AEI/ 10.13039/501100011033/FEDER, UE. 

This work was carried out on the territory of xučyun (Huichin), the ancestral and unceded land of the Chochenyo speaking Ohlone people, the successors of the sovereign Verona Band of Alameda County.

\end{acknowledgments}

\appendix

\section{Checking the impact of the sample's number density}\label{appendix:shot_noise}
Our simulated galaxy sample has a mean number density of $\bar{n}=27.3\,{\rm arcmin}^{-2}$. This is smaller than the nominal expectation for the \emph{gold} sample, $\bar{n}=40\,{\rm arcmin}^{-2}$. This means the shot noise is larger by a factor of $\sim 1.5$ (assuming Poissonian statistics). Let us now investigate at what scale the angular power spectrum of the galaxies in several of the individual redshift bins becomes dominated by shot noise, and compare our findings for the two number density scenarios. 

\begin{figure}[hbt!]    \includegraphics[width=0.9\columnwidth]{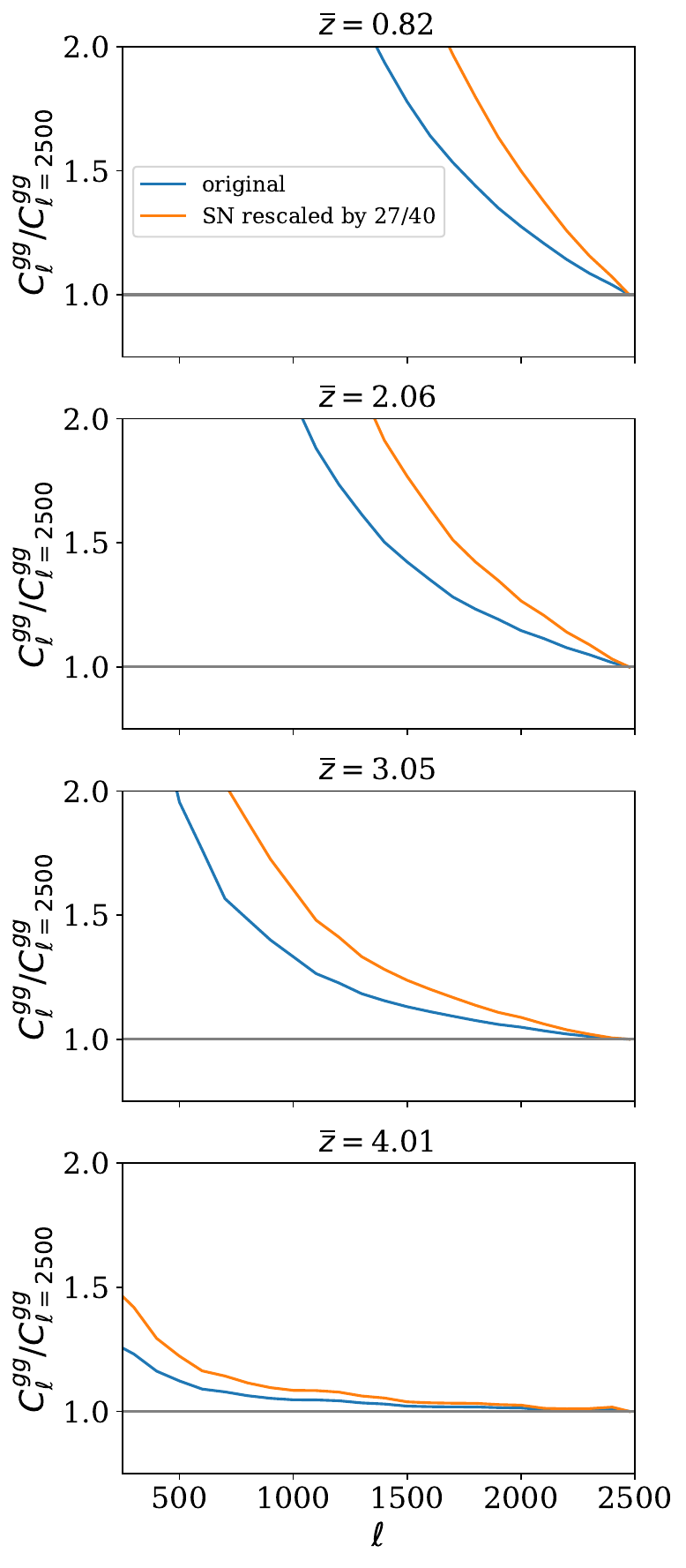}
    \caption{\label{fig:checking_shot_noise} Ratio of the galaxy angular power spectrum to its high-$\ell$ limit, intended to represent the amplitude of clustering contributions relative to shot noise as a function of angular scale. We compare results from our fiducial density (blue) to the higher density expected of the actual Rubin \emph{gold} sample (orange).}
\end{figure}

\begin{figure}[hbt!]    \includegraphics[width=0.9\columnwidth]{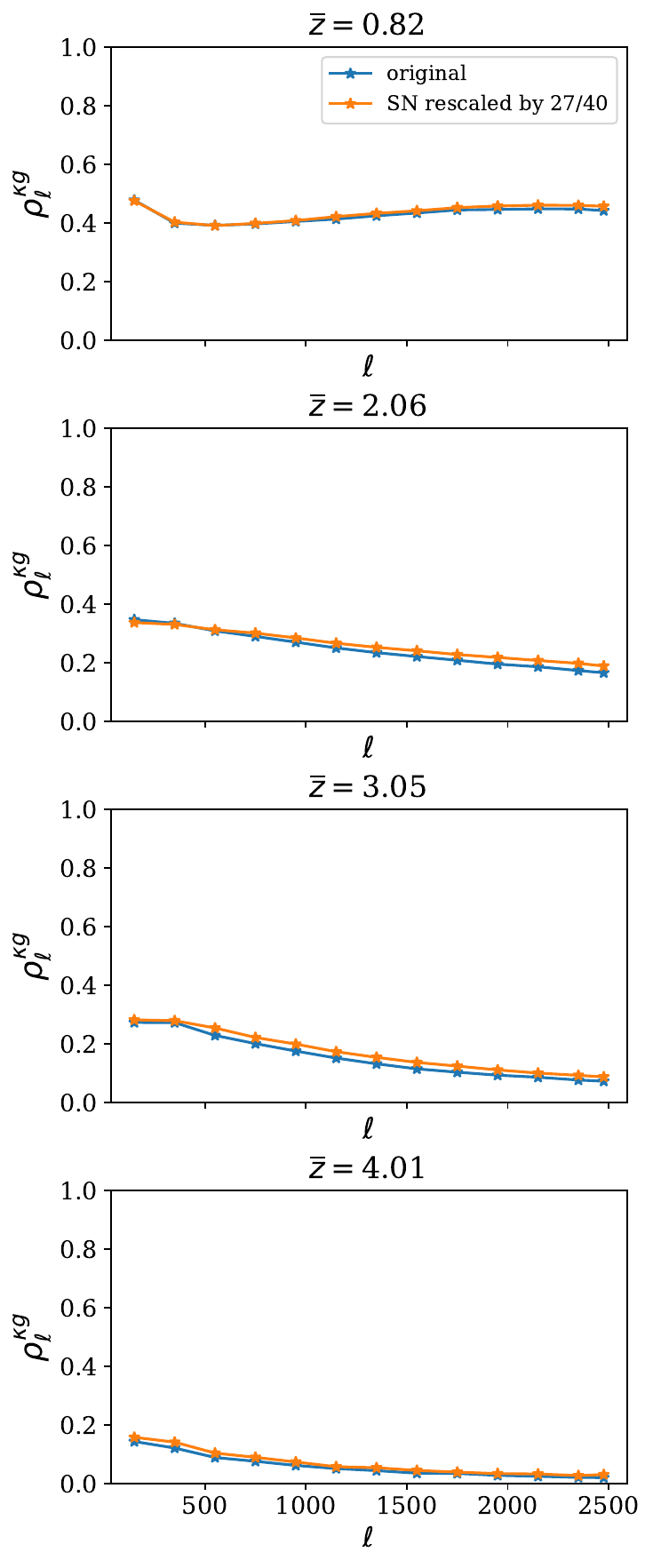}
    \caption{\label{fig:rho_after_rescaling} Impact on the tracers' cross-correlation coefficient with CMB lensing of the difference in number density between our simulated galaxies (red, solid) and that expected of the actual Rubin \emph{gold} number density (orange, dashed).}
\end{figure}

To do this, we fit a model for the angular power spectrum based around linear bias, and add an analytical shot noise component derived from the measured number density in each bin (see section~\ref{sec:forecasts}  and appendix~\ref{appendix:power_spectrum} for further details). This spectrum provides a relatively good fit to the mock data; we will take it to represent the clustering of our mocks. We will then take this model and rescale the shot noise by a factor of 27/40, producing spectra that we will take to be representative of the clustering in the actual Rubin \emph{gold} sample. In figure~\ref{fig:checking_shot_noise}, we compare the total-clustering-to-shot-noise ratio in either case. This is complemented by figure~\ref{fig:rho_after_rescaling}, where we check how these differences in shot noise level affect the samples' correlation coefficient with CMB lensing (see section~\ref{sec:theory_intro} for a definition). 

At redshifts $z\lesssim 2$, which dominate the delensing performance, differences are relatively small on the angular scales that matter most when delensing large-angular-scale $B$-modes: $200\lesssim \ell \lesssim 800$. At higher redshifts, it may be the case that our simulated samples are less correlated than the fiducial \emph{gold} sample would be. We defer a more thorough study of this regime to future work.

\section{Curved-Sky Template Delensing with Galaxy Density}
\label{appendix_curved}
In the curved-sky formalism and to first order in lensing,  a template of the lensed $B$-modes can be written as
\begin{eqnarray}
\tilde{B}^{\rm temp}_{\ell m}=\frac{(-1)^{m}}{2}\sum_{(\ell m)_{1}}\sum_{(\ell m)_{2}}\begin{pmatrix}
            \ell_{1} & \ell_{2} & \ell\\
            m_{1} & m_{2} & -m
            \end{pmatrix}\nonumber \\
\times W_{\ell_{1}\ell_{2}\ell}\tilde{E}_{(\ell m)_{2}}f_{\ell}\phi^{\rm proxy}_{(\ell m)_{1}}\,,
\label{map-level_template}
\end{eqnarray}
where $\tilde{E}$ is the lensed $E$-mode, $\phi^{\rm proxy}_{\ell m}$ is a proxy to the lensing potential, and $W$ are weights defined as:
\begin{eqnarray}
W_{\ell_{1}\ell_{2}\ell}=\frac{-i}{2}\sqrt{(2\ell+1)(2\ell_{1}+1)(2\ell_{2}+1)/4\pi}\nonumber\\
\times [\ell_{1}(\ell_{1}+1)+\ell_{2}(\ell_{2}+1)-\ell(\ell+1)]\nonumber\\
\times \left[\begin{pmatrix}
            \ell_{1} & \ell_{2} & \ell\\
            0 & -2 & 2
            \end{pmatrix}-
            \begin{pmatrix}
            \ell_{1} & \ell_{2} & \ell\\
            0 & 2 & -2
            \end{pmatrix}
            \right]\nonumber\\
            \equiv \sum_{i} W^{i}_{\ell_{1}\ell_{2}\ell},
\label{map-level_weights}
\end{eqnarray}
with
\begin{eqnarray}
W^{i}_{\ell_{1}\ell_{2}\ell}=\sqrt{(2\ell+1)(2\ell_{1}+1)(2\ell_{2}+1)/4\pi}\nonumber\\
\times \begin{pmatrix}
            \ell_{1} & \ell_{2} & \ell\\
            -s^{i}_{1} & -s^{i}_{2} & s^{i}
            \end{pmatrix}w^{i}_{\ell_{1}}w^{i}_{\ell_{2}}w^{i}_{\ell},
\label{map-level_weightsi}
\end{eqnarray}
in which the separable weights $s^{i}_{j}$ and $w^{i}_{\ell_{j}}$ are given by Table~\ref{tab:separable_weights}. The latter factorization enables a fast, position-space implementation (see, e.g.~\cite{ref:cib_delensing_biases} and the publicly-available code \texttt{csbt}\footnote{\url{https://github.com/abaleato/curved_sky_B_template}}). On the other hand,
\begin{eqnarray}
f_{\ell} = \frac{C_{\ell}^{\phi \phi^{\rm proxy}}}{{C_{\ell}^{\phi \phi, \rm proxy}}}\,,
\label{wiener_proxy}
\end{eqnarray}
is the Wiener filter for the proxy, a ratio of the cross-spectrum of the true lensing potential with the proxy over the auto-spectrum of the proxy. We assume in this work that the simulated lensed $E$-modes are noiseless such that we do not need to Wiener filter the $E$-modes. 

\begin{table}[b]
\caption{\label{tab:separable_weights}
Separable weights for the fast position-space implementation of  the lensed $B$-mode template; see, e.g.~\cite{ref:cib_delensing_biases}.
}
\begin{ruledtabular}
\begin{tabular}{c|ccccccc}
$i$&$s^{i}_{1}$&$s^{i}_{2}$&$s^{i}$&$w^{i}_{\ell_{1}}$&$w^{i}_{\ell_{2}}$&$w^{i}_{\ell}$\\
\hline
1 & 0 & 2 & 2 & $\ell_{1}(\ell_{1}+1)$ & $-1/2$ & $i$ \\
2 & 0 & 2 & 2 & $-1/2$ & $\ell_{2}(\ell_{2}+1)$ & $i$ \\
3 & 0 & 2 & 2 & 1/2 & $i$ & $\ell(\ell+1)$ \\
4 & 0 & $-2$ & $-2$ & $\ell_{1}(\ell_{1}+1)$ & 1/2 & $i$ \\
5 & 0 & $-2$ & $-2$ & 1/2 & $\ell_{2}(\ell_{2}+1)$ & $i$ \\
6 & 0 & $-2$ & $-2$ & $-1/2$ & $i$ & $\ell(\ell+1)$ \\
\end{tabular}
\end{ruledtabular}
\end{table}

\section{Modeling angular power spectra \\ of galaxies and CMB lensing}\label{appendix:power_spectrum}
Since galaxies are biased tracers of the underlying matter distribution~\cite{ref:kaiser_84}, their number (over-)density can be used as a proxy of the lensing convergence for the purpose of delensing. The expected residual lensing $B$-mode power spectrum resulting from the process is then given by equation~\eqref{eqn:full_model}, replacing $\kappa^{\rm proxy}$ with the galaxy overdensity $g$. This assumes that the $\langle \kappa \kappa g\rangle$ and $\langle \kappa g g\rangle$ bispectra are negligible --- in section~\ref{sec:results}, we check the validity of this assumption. We found that neglecting the bispectra induces percent level deviations in the residual $B$-mode power spectrum, below cosmic variance. We do not include the bispectra in this section and all results that do not depend on the $N$-body simulations.

Granted this, all we need to model the residual lensing $B$-mode power is a theoretical prediction for the galaxy auto-spectrum, $C_{\ell}^{gg}$, and cross-spectrum with CMB lensing, $C_{\ell}^{\kappa g}$, as these go into the cross-correlation coefficient in equation~\eqref{eqn:full_model}. The CMB lensing auto-spectrum is also required, but this is predicted very accurately within $\Lambda$CDM. Note that these spectra are also required when determining the tracer weights of equations~\eqref{f_weighting} and~\eqref{combine_weights}.

These angular spectra can be modeled as
\begin{eqnarray}
C_{\ell}^{ab}=4\pi\int^{\infty}_{0}\frac{dk}{k}P_{\Phi}(k)\Delta_{\ell}^{a}(k)\Delta_{\ell}^{b}(k),
\label{flat_spectrum}
\end{eqnarray}
where $P_{\Phi}(k)$ is the power spectrum of the primordial curvature perturbation at wavenumber $k$, and $\Delta_{\ell}^{a}$ and $\Delta_{\ell}^{b}$ are the transfer functions of the two tracers.

The transfer function of galaxy density is
\begin{eqnarray}
\Delta_{\ell}^{g}(k)=\int dzb(z)\frac{dN}{dz}(z)T_{\delta}(k, z)j_{\ell}(k\chi(z))\,,
\label{galaxy_transfer}
\end{eqnarray}
where $b(z)$ is a scale-independent galaxy bias given in equation~\eqref{scale-dependant_bias}, $dN/dz$ is the normalized distribution of sources in redshift, $T_{\delta}$ is the matter overdensity transfer function, and $j_{\ell}$ is the $\ell$-th order spherical Bessel function.

In addition to the result of equation~(\ref{flat_spectrum}), a Poissonian shot noise component must be added when modeling the galaxy density auto power spectrum:
\begin{eqnarray}
C_{\ell}^{gg}=4\pi\int^{\infty}_{0}\frac{dk}{k}P_{\Phi}(k)\Delta_{\ell}^{g}(k)\Delta_{\ell}^{g}(k)+N_{\ell},
\label{limber_spectrum_galaxy}
\end{eqnarray}
with
\begin{eqnarray}
N_{\ell}=f_{\rm sky}\frac{4\pi}{N_{g}},
\label{shot_noise}
\end{eqnarray}
where $f_{\rm sky}$ is the fraction of the sky covered by the observations, and $N_g$ is the total number of galaxies in the sample under scrutiny.

The transfer function for the lensing convergence is
\begin{eqnarray}
\Delta_{\ell}^{\kappa}=-\frac{\ell(\ell+1)}{2}\int^{\chi_{*}}_{0}\frac{dz}{H(z)}\frac{r(\chi_{*}-\chi)}{r(\chi)r(\chi_{*})}T_{\phi+\psi}(k,z),
\label{lensing_transfer}
\end{eqnarray}
where $\chi_{*}\equiv\chi(z_{*})$ in which $\chi(z)$ is the comoving radial distance and $z_{*}=1100$ is the redshift of the last-scattering surface, $H(z)$ is the Hubble rate, and $r(\chi)$ is the  comoving angular diameter distance.

With these expressions we can construct the correlation coefficient between galaxy density and the lensing convergence, which together with a power spectrum of the lensed $E$-modes, $C_{\ell}^{EE}$, lets us predict the lensing $B$-mode residual power spectrum, $C_{\ell}^{BB, \rm res}$, using equation~\eqref{eqn:full_model}. Note that, for a given galaxy redshift distribution, this forecast of the power spectrum residual depends on galaxy bias (cf. equation~\ref{galaxy_transfer}); in section~\ref{sec:forecasts}, we show that this dependence is rather benign, with linear bias giving a $\lesssim 5\%$ accurate description already.

\begin{figure}[hbt!]  
\includegraphics[width=0.95\columnwidth]{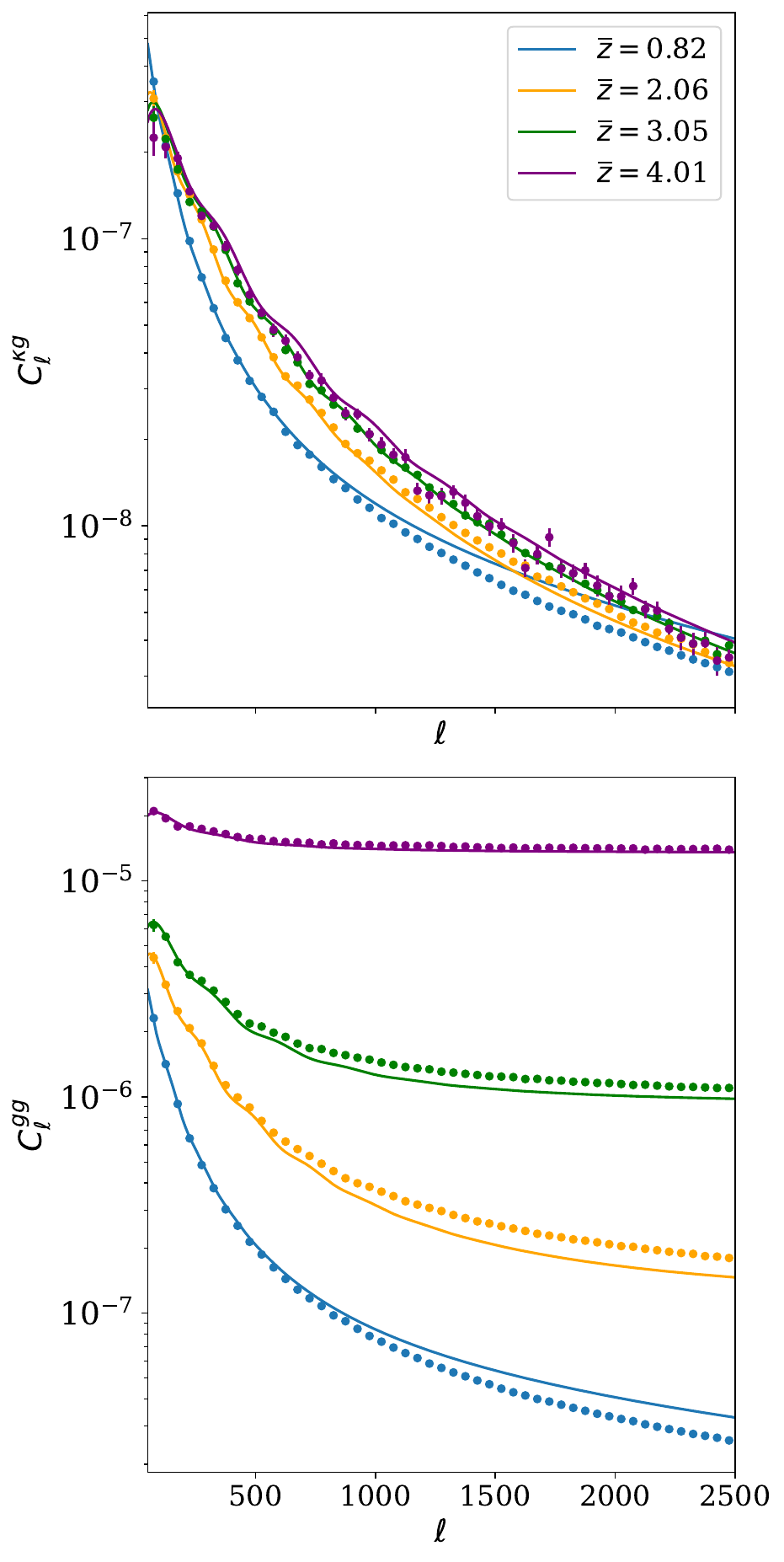}
    \caption{\label{fig:cls_fit} Angular auto-spectra (top panel) and cross-spectra with the input CMB lensing convergence (bottom panel) of our mock galaxy sample. Different colors correspond to a representative subset of the different redshift bins.}
\end{figure}

For completeness, we show in figure~\ref{fig:cls_fit} the measured angular auto- and cross-spectra with CMB lensing for a representative subsample of all of our galaxy bins. Alongside the binned raw data, we plot the best fit curves obtained using the models described in this section and the fitting procedure described in section~\ref{sec:forecasts}.

\bibliography{main}

\end{document}